\documentclass[english,journal]{ieeeaccess}
\usepackage[T1]{fontenc}
\usepackage[latin1]{inputenc}
\usepackage{amsmath}
\usepackage{graphicx}
\usepackage{amssymb}

\makeatletter

\providecommand{\tabularnewline}{\\}

 \usepackage{verbatim}

\usepackage{amsmath,amssymb,amsfonts}
\usepackage{algorithmic}

\def\BibTeX{{\rm B\kern-.05em{\sc i\kern-.025em b}\kern-.08em
    T\kern-.1667em\lower.7ex\hbox{E}\kern-.125emX}}

\usepackage{bm}

\usepackage{caption,setspace}
\usepackage{tabto}

\usepackage{babel}
\makeatother
\begin{document}
\doi{10.1109/ACCESS.2024.3397186}

\title{Optically-Transparent EM Skins for Outdoor-to-Indoor mm-Wave Wireless Communications}

\author{\uppercase{Giacomo Oliveri}\authorrefmark{1}, \IEEEmembership{Fellow, IEEE}, \uppercase{Francesco Zardi}\authorrefmark{1}, \uppercase{Giorgio Gottardi}\authorrefmark{1}, AND \uppercase{Andrea Massa}\authorrefmark{2,1,3,4}, \IEEEmembership{Fellow, IEEE}} 

\address[1]{ELEDIA Research Center (ELEDIA Research Center (ELEDIA@UniTN - University of Trento), DICAM - Department of Civil, Environmental, and Mechanical Engineering, Via Mesiano 77, 38123 Trento - Italy (e-mail: \{giacomo.oliveri, francesco.zardi, giorgio.gottardi, andrea.massa\}@unitn.it)}

\address[2]{ELEDIA Research Center (ELEDIA@UESTC - UESTC), School of Electronic Science and Engineering, University of Electronic Science and Technology of China, Chengdu 611731 - China (e-mail: andrea.massa@uestc.edu.cn)}\address[3]{ELEDIA Research Center (ELEDIA@TSINGHUA - Tsinghua University), 30 Shuangqing Rd, 100084 Haidian, Beijing - China (e-mail: andrea.massa@tsinghua.edu.cn)}

\address[4]{School of Electrical Engineering, Tel Aviv University, Tel Aviv 69978 - Israel (e-mail: andrea.massa@eng.tau.ac.il)}\tfootnote{(c) 2024 IEEE.  Personal use of this material is permitted.  Permission from IEEE must be obtained for all other uses, in any current or future media, including reprinting/republishing this material for advertising or promotional purposes, creating new collective works, for resale or redistribution to servers or lists, or reuse of any copyrighted component of this work in other works. This work benefited from the networking activities carried out within the Project SPEED (Grant No. 61721001) funded by National Science Foundation of China under the Chang-Jiang Visiting Professorship Program, the Project "National Centre for HPC, Big Data and Quantum Computing (CN HPC)" funded by the European Union - NextGenerationEU within the PNRR Program (CUP: E63C22000970007), the Project DICAM-EXC funded by the Italian Ministry of Education, Universities and Research (MUR) (Departments of Excellence 2023-2027, grant L232/2016), and the Project "AURORA - Smart Materials for Ubiquitous Energy Harvesting, Storage, and Delivery in Next Generation Sustainable Environments" funded by the Italian Ministry for Universities and Research within the PRIN-PNRR 2022 Program.}

\corresp{Corresponding author: Andrea Massa (e-mail: andrea.massa@unitn.it).}

\begin{abstract}
\noindent Optically-transparent opportunistic electromagnetic skins
(\emph{OTO-EMS}s) are proposed to enable outdoor-to-indoor (\emph{O2I})
millimeter-wave (\emph{mmW}) wireless communications with existing
windows/glass-panels. More in detail, static passive \emph{EMS}s consisting
of optically-transparent conducting patterned layers attached to standard
glass-panels are designed. Towards this end, both the phase coverage
and the optical transparency of a meshed copper-based meta-atom printed
on a non-dedicated insulated glass substrate are optimized. Successively,
the feasibility of \emph{OTO-EMS}s able to support \emph{mmW} high-efficiency
\emph{O2I} transmissions along non-Snell refraction directions is
numerically demonstrated also through full-wave simulations.
\end{abstract}

\titlepgskip=-15pt 

\maketitle

\renewcommand{\figurename}{FIGURE}

\renewcommand{\tablename}{TABLE}

\renewcommand{\refname}{REFERENCES}

\section{Introduction and Rationale\label{sec:Introduction}}

\noindent \PARstart{T}{he deployment} of wireless cellular systems
operating at millimeter wave (\emph{mmW}) frequencies has been proposed
as a suitable technological solution for the ever-increasing demand
for high data rates in mobile communications \cite{Rappaport 2017}-\cite{Rappaport 2017b}.
As a matter of fact, the availability of a wide spectrum in the \emph{mmW}
band potentially supports several Gbps of peak data rates, while still
employing relatively simple user terminals \cite{Rappaport 2017}.
However, major propagation challenges need to be addressed if outdoor-to-indoor
(\emph{O2I}) operations are of interest \cite{Rappaport 2017}, \cite{Rodriguez 2015}.
Indeed, the signal degradation caused by the building penetration
losses can prevent the establishment of a \emph{O2I} link in most
practical scenarios \cite{Rappaport 2017}, \cite{Rodriguez 2015},
\cite{Rappaport 2013}. For instance, penetration loss values exceeding
$40$ dB have been measured through outdoor glass-panels at $28$
{[}GHz{]} \cite{Rappaport 2017}, \cite{Rappaport 2013}. Similar
results have been obtained when dealing with concrete or brick walls,
as well \cite{Rappaport 2017}. To compensate for these losses, additional
indoor base stations or higher transmitting powers may be used \cite{Rappaport 2017},
\cite{Rappaport 2013}, but such strategies would imply a non-trivial
increase in the network complexity, the operational costs, and the
energy consumption.%
\begin{figure}
\begin{center}\begin{tabular}{c}
\includegraphics[%
  width=0.95\columnwidth,
  keepaspectratio]{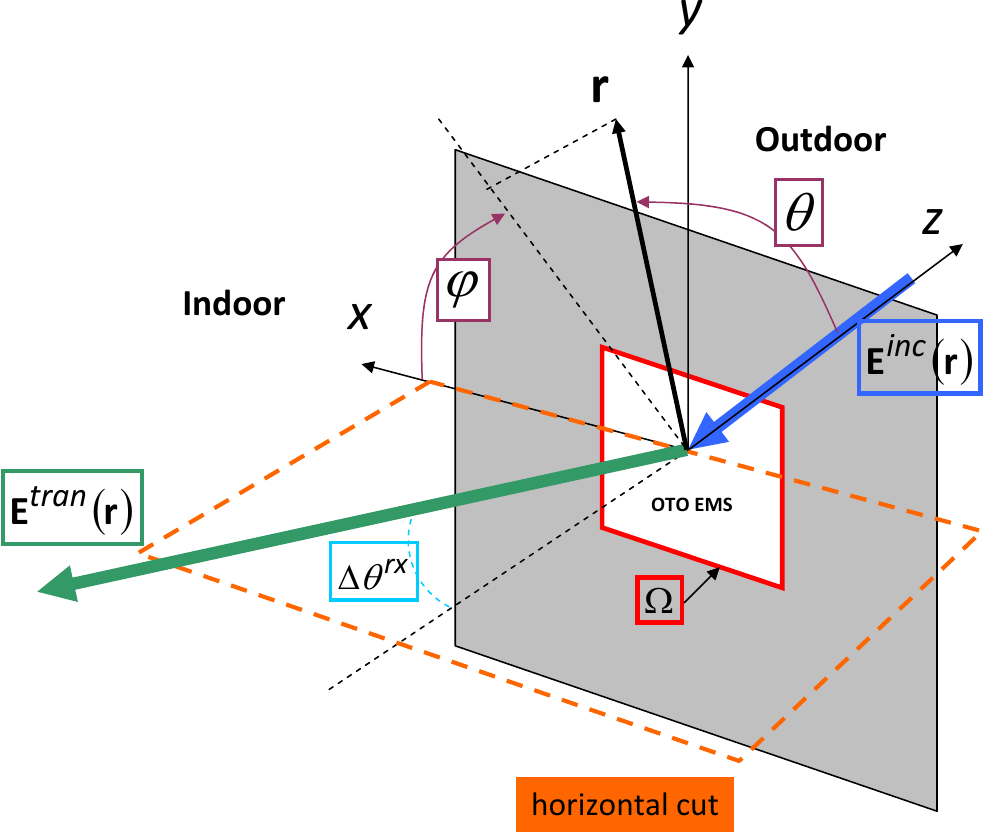}\tabularnewline
(\emph{a})\tabularnewline
\includegraphics[%
  width=0.95\columnwidth,
  keepaspectratio]{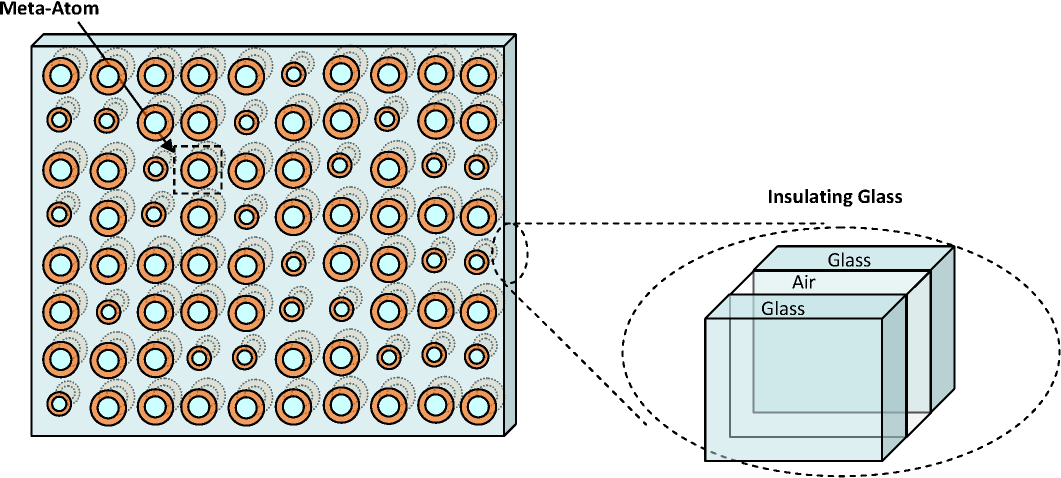}\tabularnewline
(\emph{b})\tabularnewline
\tabularnewline
\tabularnewline
\end{tabular}\end{center}

\caption{Sketch of (\emph{a}) the \emph{O2I} scenario at hand and (\emph{b})
the \emph{OTO-EMS} architecture.}
\end{figure}

\noindent An intriguing alternative is the inclusion in the building
walls of passive field manipulating devices to {}``route'' the electromagnetic
(\emph{EM}) propagation according to the desired \emph{O2I} paths
{[}Fig. 1(\emph{a}){]}. Such an idea stems from the recent methodological
and technological advances in surface \emph{EM} and artificial 2D
material engineering \cite{Yang 2019} at the basis of the Smart \emph{EM}
Environment (\emph{SEME}) paradigm \cite{Massa 2021}-\cite{Huang 2019}.
More specifically, a planar \emph{EMS} capable of tailoring the refracted
field may be designed so that the penetration loss through its surface
is minimized and, additionally, the transmitted field propagates along
a non-Snell direction \cite{Yang 2019}. A {}``traditional'' design
of such a transmitting \emph{EMS} would typically require several
conducting layers separated by a combination of air, foam, and/or
dedicated substrates \cite{Yang 2019}. Moreover, and besides the
non-negligible costs, substituting a portion of a wall or a window
with such an \emph{EMS} would be impractical because of both the visual
impact and the structural/thermal insulation limitations.

\noindent To make sustainable \emph{O2I} wireless communications at
\emph{mmW} frequencies, an opportunistic implementation is proposed
in this paper. More specifically, the concept of optically-transparent
opportunistic \emph{EMS} (\emph{OTO-EMS}) is introduced. An \emph{OTO-EMS}
is a static passive \emph{EMS} that consists of one or more conducting,
but optically-transparent, patterned layers which are attached using
an optical clear adhesive (\emph{OCA})%
\footnote{\emph{OCA}s are typically based on acrylate chemistry, and exhibit
electrical properties similar to glass \cite{Abrahamson 2023}. Several
\emph{OCA} technologies and fabrication processes are commercially
available \cite{Abrahamson 2023}.%
} to an \emph{existing} glass window, this latter acting as an equivalent
\emph{EMS} support/substrate {[}Fig. 1(\emph{b}){]}. Accordingly,
an \emph{OTO-EMS} can be seamlessly installed on any existing window
to establish a reliable \emph{O2I} link at \emph{mmW} frequencies
by avoiding any visual/structural impact or thermal/acoustic insulation
issues of traditional implementations.

\noindent Hereafter, the design of \emph{OTO-EMS}s will be carried
out by addressing the following implementation challenges:

\begin{itemize}
\item \noindent the design of a meta-atom%
\footnote{\noindent In the following, {}``meta-atom'' and {}``unit cell''
will be used as synonyms to identify the elementary \emph{EMS} element,
likewise in the recent literature on \emph{SEME} \cite{Yang 2019},
\cite{Yang 2022}, \cite{Oliveri 2021c}-\cite{Oliveri 2023b}.%
} that, unlike state-of-the-art layouts for transmitting skins \cite{Yang 2019},
combines a set of patterned transparent conductive layers and a non-dedicated
glass-based substrate to yield a suitable phase coverage and a \emph{mmW}
transparency;
\item the synthesis of finite \emph{mmW} \emph{OTO-EMS}s featuring anomalous
wave manipulation capabilities (e.g., non-Snell refraction) and high
aperture efficiencies.
\end{itemize}
\noindent To achieve these goals, a customized implementation of the
System-by-Design (\emph{SbD}) paradigm \cite{Massa 2022}, recently
demonstrated for standard reflecting \emph{EMS} engineering \cite{Oliveri 2021c}-\cite{Oliveri 2023b},
is used. More specifically, the meta-atom design process is based
on a parametric approach featuring (\emph{i}) an elementary geometrical
layout and (\emph{ii}) the exploitation of a standard commercial \emph{insulating
glass} (\emph{IG}) as the \emph{EMS} substrate, which consists of
two glass panes separated by an air-filled region {[}Fig. 1(\emph{b}){]}.
Furthermore, the optically-transparent conductors to realize the \emph{EMS}
patterning are realized with the \emph{copper mesh} concept \cite{Sharma 2019}-\cite{Silva 2021b}
because of its better suitability for \emph{mmW} applications and
the relatively inexpensive implementation if compared to transparent
conducting oxides \cite{Silva 2021b} or liquid crystals \cite{Kim 2022}.
The multi-atom structure composing the \emph{EMS} is then optimized
to fit macro-scale performance goals expressed in terms of transmitted
field and optical transparency.

\noindent According to such a description and to the best of the authors'
knowledge, the main novelties of this work with respect to the state
of the art include (\emph{i}) the use at \emph{mmW} frequencies of
static passive transmitting \emph{EMS}s suitable as a retro-fitting
option for existing windows/glass-panels; (\emph{ii}) the opportunistic
implementation of \emph{EMS}s by exploiting the standard insulating
glass as \emph{EMS} substrate and featuring transparent conductors
for the metallic patterning; (\emph{iii}) the assessment of the effectiveness
of the proposed \emph{O2I} solution in terms of wave manipulation
properties and achievable improvements with respect to standard glass-panels
both in standard and non-Snell directions; (\emph{iv}) the customization
of the system-by-design synthesis method \cite{Massa 2022} to the
case of transmitting \emph{EMS}s, while state-of-the-art implementations
refer to reflecting surfaces \cite{Oliveri 2021c}-\cite{Oliveri 2023b}.

\noindent The outline of the paper is as follows. The design problem
at hand is formulated in Sect. II, where the \emph{OTO-EMS} concept
is introduced, as well. Section III reports a set of representative
numerical results to show the features and to assess, also through
full-wave simulations, the potentialities of \emph{OTO-EMS}s as retro-fitting
options of existing windows/glass-panels to support \emph{O2I} \emph{mmW}
wireless communications. Finally, some conclusions are drawn (Sect.
IV).

\section{\noindent Mathematical Formulation\label{sec:Problem-Formulation} }

\noindent Let us consider the \emph{O2I} wireless communications scenario
in Fig. 1(\emph{a}) where an outdoor source illuminates an \emph{OTO-EMS},
which occupies a region $\Omega$ of a glass window, with an incident
time-harmonic electromagnetic field $\mathbf{E}^{inc}$. This latter
is locally modeled as a plane wave with wave vector $\mathbf{k}^{inc}$
and incidence direction $\left(\theta^{inc},\varphi^{inc}\right)$.
The field $\mathbf{E}^{tran}$ transmitted through the region $\Omega$
in a point $\mathbf{r}$ of \emph{local} coordinates $\left(x,y,z\right)$
depends on the vector $\mathcal{D}$ {[}i.e., $\mathbf{E}^{tran}=\mathbf{E}^{tran}\left(\mathbf{r};\mathcal{D}\right)${]}
of the geometrical/physical descriptors of the \emph{EMS} \begin{equation}
\mathcal{D}\triangleq\left\{ \underline{d}_{pq};\, p=1,...,P;\, q=1,...,Q\right\} \label{eq:}\end{equation}
whose $L$-sized ($p$, $q$)-th ($p=1,...,P$; $q=1,...,Q$) entry,
$\underline{d}_{pq}\triangleq\left\{ d_{pq}^{\left(l\right)};\, l=1,...,L\right\} $,
contains the features of the corresponding meta-atom including the
\emph{copper mesh} descriptors. 

\noindent Under the assumption that the \emph{O2I} transmission from
other building portions (e.g., walls) is negligible, the synthesis
of an \emph{OTO-EMS} for establishing a reliable \emph{O2I} wireless
link can be formulated as the identification of the \emph{EMS} descriptors
$\mathcal{D}$ such that $\Phi\left[\mathbf{E}^{tran}\left(\mathbf{r};\mathcal{D}\right)\right]$
is minimized\begin{equation}
\mathcal{D}^{opt}\triangleq\arg\left\{ \min_{\mathcal{D}}\left[\Phi\left[\mathbf{E}^{tran}\left(\mathbf{r};\mathcal{D}\right)\right]\right]\right\} \label{eq:Dopt}\end{equation}

\noindent where $\Phi\left[\mathbf{E}^{tran}\left(\mathbf{r};\mathcal{D}\right)\right]$
is a cost function that encodes the macro-scale radiation objectives
defined on the total transmitted pattern. 

\noindent Regardless of the definition of $\Phi$, a reliable methodology
to evaluate $\mathbf{E}^{tran}\left(\mathbf{r};\mathcal{D}\right)$
is necessary. Towards this end, the full-wave modeling of the \emph{OTO-EMS}
could be adopted, but such an approach is practically infeasible due
to the associated computational costs and even more if it had to be
iteratively repeated for every guess of $\mathcal{D}$ as generally
required in \emph{SbD}-based design processes \cite{Massa 2022}.

\noindent Consequently, an alternative semi-analytic state-of-the-art
approach that leverages on the Love's equivalence principle and the
homogenization of the surface currents \cite{Yang 2019}, \cite{Oliveri 2021c}-\cite{Oliveri 2023b}\cite{Balanis 1989}-\cite{Pearson 2022}
is adopted in the following. More specifically, the field transmitted
through the region $\Omega$ in the far field is computed as follows
\cite{Oliveri 2023}, \cite{Mencagli 2020}

\noindent \begin{equation}
\begin{array}{l}
\mathbf{E}^{tran}\left(\mathbf{r};\mathcal{D}\right)\approx\frac{jk_{0}}{4\pi}\frac{\exp\left(-jk_{0}\left|\mathbf{r}\right|\right)}{\left|\mathbf{r}\right|}\times\\
\sum_{p=1}^{P}\sum_{q=1}^{Q}\widehat{\mathbf{r}}'\times\left[\eta_{0}\widehat{\mathbf{r}}'\times\mathbf{J}_{pq}^{e}\left(\underline{d}_{pq}\right)+\mathbf{J}_{pq}^{m}\left(\underline{d}_{pq}\right)\right]\times\\
\int_{\Omega_{pq}}\exp\left(jk_{0}\widehat{\mathbf{r}}\cdot\mathbf{r}'\right)\mathrm{d}\mathbf{r}'\end{array}\label{eq:field-finale}\end{equation}
where $\widehat{\mathbf{r}}'\triangleq\frac{\mathbf{r}'}{\left|\mathbf{r}'\right|}$,
$k_{0}$ and $\eta_{0}$ are the free-space wave-number and impedance,
respectively,\begin{equation}
\left\{ \begin{array}{l}
\mathbf{J}_{pq}^{e}\left(\underline{d}_{pq}\right)\approx\frac{1}{\eta_{0}}\widehat{\mathbf{z}}\times\mathbf{k}^{inc}\left(\mathbf{r}_{pq}\right)\times\left[\overline{\overline{T}}_{pq}\left(\underline{d}_{pq}\right)\cdot\mathbf{E}^{inc}\left(\mathbf{r}_{pq}\right)\right]\\
\mathbf{J}_{pq}^{m}\left(\underline{d}_{pq}\right)\approx\left[\overline{\overline{T}}_{pq}\left(\underline{d}_{pq}\right)\cdot\mathbf{E}^{inc}\left(\mathbf{r}_{pq}\right)\right]\times\widehat{\mathbf{z}}\end{array}\right.\label{eq:coefficienti correnti}\end{equation}
are the electric/magnetic surface current coefficients in the $pq$-th
meta-atom, $\mathbf{r}_{pq}$ is the barycenter of the $pq$-th meta-atom
area, \begin{equation}
\overline{\overline{T}}_{pq}\left(\underline{d}_{pq}\right)=\left[\begin{array}{cc}
\left.T_{pq}\left(\underline{d}_{pq}\right)\right\rfloor _{TE} & \left.T_{pq}\left(\underline{d}_{pq}\right)\right\rfloor _{TE/TM}\\
\left.T_{pq}\left(\underline{d}_{pq}\right)\right\rfloor _{TM/TE} & \left.T_{pq}\left(\underline{d}_{pq}\right)\right\rfloor _{TM}\end{array}\right]\label{eq:}\end{equation}
 is the local complex transmission tensor \cite{Cuesta 2018} (also
labeled as local dyadic transmission coefficient \cite{Mencagli 2020})
in the $pq$-th meta-atom, under the local periodicity approximation
\cite{Oliveri 2021c}, with support $\Omega_{pq}$ ($\sum_{p=1}^{P}\sum_{q=1}^{Q}\Omega_{pq}\triangleq\Omega$)
of area $\Delta^{2}$, $\Delta$ being the meta-atom lattice periodicity
so that the \emph{EMS} has a total area of $\mathcal{L}_{x}\times\mathcal{L}_{y}=\left(P\times\Delta\right)\times\left(Q\times\Delta\right)$. 

\noindent According to (\ref{eq:field-finale}), the macro-scale behaviour
of an \emph{OTO-EMS} depends on the magnitude and the phase of the
$\overline{\overline{T}}_{pq}\left(\underline{d}_{pq}\right)$ entries,
which are controlled by the user-defined micro-scale descriptors $\mathcal{D}$.
However, unlike the synthesis of reflective \emph{EMS}s \cite{Oliveri 2021c}-\cite{Oliveri 2023b}\cite{Oliveri 2023},
additional difficulties arise for the \emph{EMS} designer owing to
the phase/magnitude limits of such entries \cite{Abdelrahman 2014}.
As a matter of fact, while a single-layer patterning is sufficient
to yield a full $360^{\circ}$ phase coverage in reflecting \emph{EMS}s
\cite{Yang 2019}, a transmitting \emph{EMS} must theoretically include
at least three layers of conducting material to do the same with a
minimum of $50\%$ of the transmission efficiency \cite{Abdelrahman 2014},
unless Huygens' designs are considered \cite{Lian 2021}-\cite{Su 2023}.
On the other hand, the requirement of easy integration in existing
windows of our \emph{OTO-EMS} forces the meta-atom architecture to
a two-layer metallic structure {[}Fig. 1(\emph{b}){]} with an intrinsic
limitation in terms of transmission coefficient control.

\noindent According to the above formulation, it turns out that (\emph{i})
$\mathbf{E}^{tran}$ is determined by the surface-current coefficients
\{$\mathbf{J}_{pq}^{w}$ ($w=\left\{ e,m\right\} $); $p=1,...,P$;
$q=1,...,Q$\} according to the linear equation (\ref{eq:field-finale}),
while (\emph{ii}) these latter, which satisfy (\ref{eq:coefficienti correnti}),
depend on the entries of $\overline{\overline{T}}_{pq}$ that are
non-linearly related to the meta-atom descriptors $\underline{d}_{pq}$
{[}i.e., $\overline{\overline{T}}_{pq}=\overline{\overline{T}}\left(\underline{d}_{pq}\right)${]}.
Consequently and analogously to the guidelines adopted in the design
of reflecting \emph{EMS}s \cite{Oliveri 2021c}, \cite{Oliveri 2022},
the \emph{OTO-EMS} synthesis problem is split into two sub-problems:
(1) the \emph{OTO} meta-atom design and (2) the \emph{OTO-EMS} layout
synthesis.%
\begin{figure}
\begin{center}\begin{tabular}{c}
\includegraphics[%
  width=0.70\columnwidth,
  keepaspectratio]{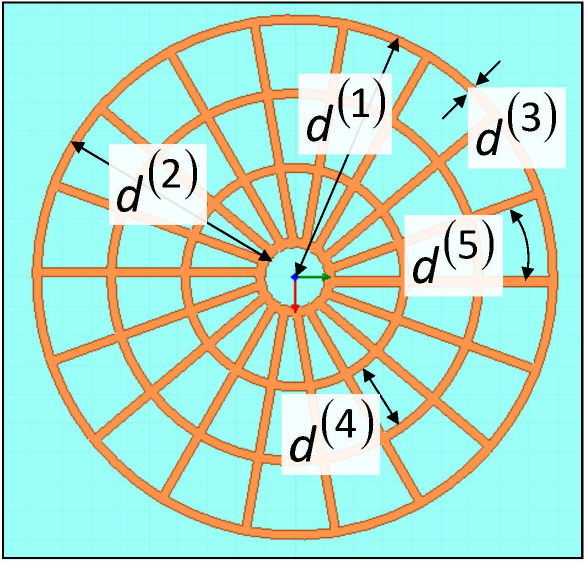}\tabularnewline
(\emph{a})\tabularnewline
\includegraphics[%
  width=0.30\columnwidth,
  keepaspectratio,
  angle=90]{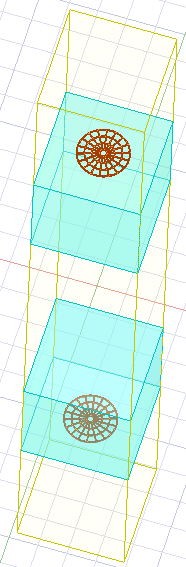}\tabularnewline
(\emph{b})\tabularnewline
\tabularnewline
\tabularnewline
\end{tabular}\end{center}

\caption{\emph{OTO Meta-Atom Design} - Picture of (\emph{a}) the parametric
representation of the unit cell and (\emph{b}) the corresponding HFSS
3D model.}
\end{figure}

\subsection{\noindent \emph{OTO} Meta-Atom Design\label{sub:OTO-EMS-Meta-Atom-Synthesis}}

As for the former sub-problem (1), it is stated as follows

\begin{quotation}
\noindent \textbf{\emph{\underbar{OTO Meta-Atom Design Problem}}}
- Design a meta-atom structure and $\Psi^{opt}$ so that $\Phi_{atom}\left(\Psi\right)=\left\{ \sum_{i=TE,TM}\left[\alpha_{i}^{PC}\Phi_{i}^{PC}\left(\Psi\right)+\alpha_{i}^{MAG}\Phi_{i}^{MAG}\left(\Psi\right)\right]+\right.$
$\left.\alpha^{OT}\Phi^{OT}\left(\Psi\right)\right\} ^{-1}$ is minimized
\end{quotation}
\noindent where $\Psi$ is the meta-atom feasibility set ($\Psi\triangleq\left\{ \underline{d}\,:\, d_{\min}^{\left(l\right)}\leq d^{\left(l\right)}\leq d_{\max}^{\left(l\right)};\, l=1,...,L\right\} $),
$\Phi_{i}^{PC}$ is the phase coverage of the unit cell on the $i$-th
($i=\left\{ TE,TM\right\} $) field component \cite{Yang 2019} ($\Phi_{i}^{PC}\left(\Psi\right)\triangleq\max_{\underline{d}\in\Psi}\angle T_{i}\left(\underline{d}\right)-$
$\min_{\underline{d}\in\Psi}\angle T_{i}\left(\underline{d}\right)$),
$\Phi_{i}^{MAG}$ is the worst-case transmittance magnitude coefficient
on the $i$-th field component ($\Phi_{i}^{MAG}\left(\Psi\right)=\min_{\underline{d}\in\Psi}\left[\left|T_{i}\left(\underline{d}\right)\right|\right]$),
$\alpha_{i}^{PC}$ and $\alpha_{i}^{MAG}$ being the corresponding
weights, while $\Phi^{OT}$ and $\alpha^{OT}$ are the optical transparency
and the associated weight, respectively. It is worth remarking that
both \emph{TE} and \emph{TM} modes have been included in the definition
of $\Phi_{atom}\left(\Psi\right)$ since their effect has been considered
in the subsequent results. Moreover, the definition of $\Phi^{OT}$
depends on both the meta-atom geometry (i.e., patterning layout) and
the geometry of the conducting mesh. In this paper, it is computed
as $\Phi^{OT}\left(\Psi\right)\triangleq\min_{\underline{d}\in\Psi}\left[\mathcal{T}_{atom}^{OTO}\left(\underline{d}\right)\right]$
where $\mathcal{T}_{atom}^{OTO}\left(\underline{d}\right)$ is the
overall meta-atom optical transmittance \cite{Silva 2021b}.

\noindent The first step in solving the \emph{OTO Meta-Atom Design
Problem} is the choice of the meta-atom model and the corresponding
unit cell descriptors $\underline{d}$. Towards this end, one should
notice that, unlike other \emph{EMS} design problems \cite{Yang 2019},
the addressed scenario is strongly constrained since the amount of
conducting material of the \emph{copper mesh} patterning in the unit
cell area has to be minimized for maximizing the optical transparency
$\mathcal{T}$, while still allowing a suitable \emph{mmW} field transmission
control. Moreover, the layout of the unit cell has to enable a direct
installation on commercial \emph{IG} windows with non-customized thicknesses
and profiles without the need for additional substrates/supports and/or
any further mechanical operation such as vias drilling. On the other
hand, as in antenna synthesis \cite{Lin 2005}, \cite{Johnson 1999},
different codings of the structure at hand, including parametric and
non-parametric descriptions \cite{Lin 2005}, \cite{Johnson 1999},
imply different solution approaches.

\noindent According to such considerations, a canonical meta-atom
geometry is chosen and a parametric coding of such a reference structure
is adopted \cite{Yang 2019}, \cite{Oliveri 2021c}. The main motivation
is that of demonstrating the feasibility of the \emph{OTO-EMS} concept
without the need for advanced meta-atom shaping and/or complex coding.
More in detail, the dual-layer circular ring meta-atom \cite{Yang 2019}
in Fig. 2 ($L=5$) is adopted. In this case, the conductive copper
mesh is implemented according to a conformal geometry combining circular
and radial wires and it is printed on both sides of the glass {[}Fig.
2(\emph{a}){]}. The meta-atom descriptors are the outer ring radius
$d^{\left(1\right)}$, the overall ring width $d^{\left(2\right)}$,
the mesh wire radius $d^{\left(3\right)}$, the mesh radial gap $d^{\left(4\right)}$,
and the mesh angular gap $d^{\left(5\right)}$ {[}Fig. 2(\emph{a}){]}.
All the remaining features (i.e., the glass/air layer thickness and
material properties) are user-defined constants. Moreover, each patterned
layer is assumed to be attached to the outer surface of the glass
panel {[}Fig. 2(\emph{b}){]} to make it suitable as a retro-fitting
option for existing windows.

\noindent Thanks to these choices, the atom is expected to have good
performance stability whatever the incident polarization \cite{Yang 2019},
while yielding a low atom fill factor $\mathcal{F}_{atom}^{OTO}$
(i.e., the ratio between the conductive area and the meta-atom support),
hence resulting in high optical transparency \cite{Silva 2021b}.
Numerically, it turns out that\begin{equation}
\mathcal{F}_{atom}^{OTO}\approx\frac{\pi d^{\left(2\right)}\left[2d^{\left(1\right)}-d^{\left(2\right)}\right]}{\Delta^{2}}\mathcal{F}_{mesh},\label{eq:ratio_area}\end{equation}
$\mathcal{F}_{mesh}\triangleq\frac{d^{\left(3\right)}}{d^{\left(3\right)}+d^{\left(4\right)}}$
being the mesh fill factor \cite{Silva 2021b} that corresponds to
an atom optical transmittance of $\mathcal{T}_{atom}^{OTO}=\left(1-\mathcal{F}_{atom}^{OTO}\right)^{4}$
under the assumption of perfect transparency of the glass \cite{Silva 2021b}.
For comparison purposes, let us notice that $\mathcal{F}_{atom}^{OTO}$
is much lower than the fill-factor of standard \emph{EMS} meta-atoms
such as printed square patches \cite{Oliveri 2021c}, $\mathcal{F}_{atom}^{Patch}=\frac{\left[d^{patch}\right]^{2}}{\Delta^{2}}$,
$d^{patch}$ being the patch side.

\noindent As far as the design procedure is concerned, thanks to the
reduced number of descriptors of the meta-atom {[}$L=5$ - Fig. 2(\emph{a}){]},
the responses of several guess meta-atoms belonging to different feasibility
sets are full-wave simulated with Ansys HFSS \cite{HFSS 2021} {[}Fig.
2(\emph{b}){]} until $\Phi_{atom}\left(\Psi\right)\leq\eta_{atom}$,
$\eta_{atom}$ being a user-chosen threshold.

\subsection{\noindent \emph{OTO-EMS} Layout Synthesis\label{sub:OTO-EMS-Layout-Synthesis}}

Concerning the sub-problem (2), it can be formulated as follows

\begin{quotation}
\noindent \textbf{\emph{\underbar{OTO-EMS Layout Synthesis Problem}}}
- Given $\mathbf{E}^{inc}\left(\mathbf{r}\right)$, find $\mathbf{J}_{opt}^{w}\left(\mathbf{r};\mathcal{D}\right)$
($w=\left\{ e,m\right\} $) and the corresponding $\mathcal{D}^{opt}$
such that $\Phi\left[\mathbf{E}^{tran}\left(\mathbf{r};\mathcal{D}\right)\right]$
is minimized
\end{quotation}
\noindent where $\mathbf{E}^{tran}$ is semi-analytically computed
as in (\ref{eq:field-finale}) starting from (\ref{eq:coefficienti correnti})
and using the values of the local transmission tensor $\overline{\overline{T}}$
in correspondence with the \emph{OTO} meta-atom feasibility set $\Psi^{opt}$
deduced in sub-problem (1) {[}i.e., $\overline{\overline{T}}=\overline{\overline{T}}\left(\underline{d}_{pq}\right)$,
$\underline{d}_{pq}\in\Psi^{opt}${]}.

\noindent It is worthwhile to point out that the proposed \emph{OTO-EMS}
synthesis approach, which separates the meta-atom design {[}\emph{Sub-Problem}
(1){]} from the \emph{EMS} layout synthesis {[}Sub-Problem (2){]},
allows one to deal with a wide variety of different problems, each
being formulated by re-defining the cost function $\Phi$, without
the need to re-engineer the unit cell structure.

\noindent The synthesis of an \emph{OTO-EMS} layout fitting a macro-scale
radiation objective requires the explicit definition of the associated
cost function $\Phi$. Analogously to \cite{Oliveri 2023b}, this
paper considers the power-maximization towards a receiver located
in a position $\mathbf{r}^{rx}$. Therefore, the cost function $\Phi$
is given by\begin{equation}
\Phi\left[\mathbf{E}^{tran}\left(\mathbf{r};\mathcal{D}\right)\right]\triangleq\frac{1}{\left|\mathbf{E}^{tran}\left(\mathbf{r}^{rx};\mathcal{D}\right)\right|}.\label{eq:macro-scale objective}\end{equation}
In fact, from the physical perspective, the minimization of $\Phi\left[\mathbf{E}^{tran}\left(\mathbf{r};\mathcal{D}\right)\right]$
implies to the maximization of $\left|\mathbf{E}^{tran}\left(\mathbf{r};\mathcal{D}\right)\right|$
in the $\mathbf{r}=\mathbf{r}^{rx}$ location, which corresponds to
the realization of a collimated transmitted beam towards the receiver.
Consequently and likewise state-of-the-art approaches dealing with
reflecting \emph{EMS}s \cite{Oliveri 2021c}-\cite{Oliveri 2023b},
the definition of the micro-scale descriptors $\mathcal{D}^{opt}$
is determined by means of a two-step process where (\emph{a}) the
ideal surface currents $\widetilde{\mathbf{J}}^{w}\left(\mathbf{r}\right)$
($w=\left\{ e,m\right\} $) that minimize (\ref{eq:macro-scale objective})
are firstly deduced, then (\emph{b}) $\mathcal{D}^{opt}$ is computed
by solving the following current-matching problem\begin{equation}
\mathcal{D}^{opt}=\arg\left\{ \min_{\mathcal{D}}\left[\Upsilon\left(\mathcal{D}\right)\right]\right\} \label{eq:}\end{equation}
where $\Upsilon\left(\mathcal{D}\right)\triangleq\left\Vert \mathbf{J}^{w}\left(\mathbf{r};\mathcal{D}\right)-\widetilde{\mathbf{J}}^{w}\left(\mathbf{r}\right)\right\Vert ^{2}$.

\noindent In particular, Step (\emph{a}) is addressed with the phase-conjugation
method \cite{Oliveri 2023b}, \cite{Mailloux 2005}-\cite{Pon 1964}
to yield \begin{equation}
\arg\left\{ \widetilde{J}_{a}\right\} _{pq}^{w}=-\arg\left[\int_{\Omega_{pq}}\exp\left(jk_{0}\widehat{\mathbf{r}}^{rx}\cdot\mathbf{r}'\right)\mathrm{d}\mathbf{r}'\right]\label{eq:current design equation}\end{equation}
 ($w=\left\{ e,m\right\} $, $a=\left\{ x,y\right\} $, $p=1,...,P$,
$q=1,...,Q$) where $\widehat{\mathbf{r}}^{rx}=\left\{ \sin\theta^{rx}\cos\varphi^{rx},\sin\theta^{rx}\sin\varphi^{rx},\cos\theta^{rx}\right\} $,
$\left(\theta^{rx},\varphi^{rx}\right)$ being the receiver direction.
Such a method guarantees that all current terms are added in-phase
at the focusing point so that (\ref{eq:macro-scale objective}) is
intrinsically minimized. %
\begin{figure}
\begin{center}\includegraphics[%
  width=0.95\columnwidth,
  keepaspectratio]{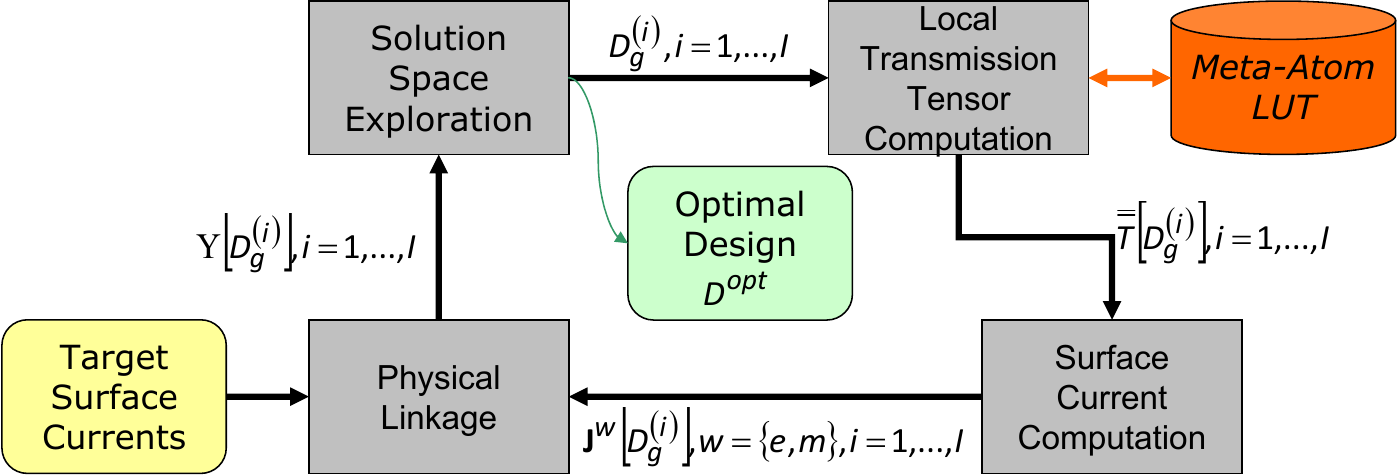}\end{center}

\caption{Flowchart of the customized SbD approach for the computation of $\mathcal{D}^{opt}$.}
\end{figure}

\noindent Step (\emph{b}) is dealt with a customized version of the
\emph{SbD} strategy \cite{Massa 2022}-\cite{Oliveri 2023b} that
generates a succession of $G$ iterations to identify $\mathcal{D}^{opt}$
(Fig. 3) \cite{Massa 2022}. At each $g$-th ($g=1,...,G$) iteration,
a set of $I$ guess \emph{EMS} configurations, \{ $\mathcal{D}_{g}^{\left(i\right)}\triangleq\left\{ \left.\underline{d}_{pq}\right\rfloor _{g}^{\left(i\right)}\in\Psi^{opt};\, p=1,...,P,\, q=1,...,Q\right\} $;
$i=1,...,I$\} are generated by combining (\emph{a}) a \emph{solution-space-exploration}
block, which is implemented in a in-house coded software according
to the particle swarm paradigm \cite{Rocca 2009}; (\emph{b}) a \emph{local
transmission tensor look-up table} (\emph{LUT}) block, which stores
the non-linear meta-atom response $\overline{\overline{T}}\left(\underline{d}\right)$,
$\underline{d}\in\Psi^{opt}$, computed through the full-wave simulation
of its numerical model in Ansys HFSS {[}Fig. 2(\emph{b}){]}; (\emph{c})
a \emph{surface electric/magnetic current computation} block, which
implements (\ref{eq:coefficienti correnti}), and (\emph{d}) a \emph{physical
linkage} block, which is responsible for the computation of $\Upsilon\left(\mathcal{D}\right)$.
Accordingly, the procedure is not directly interfaced with \emph{Ansys
HFSS}, but rather exploits the outcomes of its simulations as stored
in the \emph{LUT}. The entire process is repeated until a maximum
number of steps (i.e., $g=G$) or a stationarity condition on the
minimization of the cost function is reached \cite{Massa 2022}-\cite{Oliveri 2023b},
\cite{Rocca 2009}. The optimal \emph{OTO-EMS} descriptors are then
set as follows\begin{equation}
\mathcal{D}^{opt}=\arg\left\{ \min_{g=1,...,G;i=1,...,I}\left[\min\Upsilon\left(\mathcal{D}_{g}^{\left(i\right)}\right)\right]\right\} .\label{eq:}\end{equation}

\section{\noindent Numerical Results\label{sec:Results}}

\noindent The objective of this section is twofold. On the one hand,
it is aimed at proving the feasibility of an \emph{OTO} meta-atom
that enables advanced \emph{mmW}-frequency wave manipulation properties,
while keeping good optical transparency (Sect. III.A). On the other
hand, it is devoted to show the features of \emph{OTO-EMS}s and to
assess their performance in comparison with non-patterned glasses
as well as empty windows (Sect. III.B).

\noindent Unless otherwise specified, the \emph{OTO-EMS}s designs
have been carried out at $f_{0}=26$ {[}GHz{]} by considering a $\left(\theta^{inc},\varphi^{inc}\right)=\left(0,0\right)$
{[}deg{]} incident $\varphi$-polarized wave that illuminates the
\emph{EMS} with a $1$ {[}V/m{]} field magnitude. The standard $4-10-4$
\emph{IG} has been assumed as benchmark substrate. It consists of
two glass panes of thickness $\tau_{glass}=4$ {[}mm{]}, relative
permittivity $\varepsilon_{r}=5.5$ and loss tangent $\tan\delta=3.0\times10^{-2}$,
which are separated by an air-filled space $\tau_{air}=10$ {[}mm{]}
thick (with unitary relative permittivity), so that the total thickness
is $\tau_{IG}=18$ {[}mm{]} (i.e., $\tau_{IG}\approx1.56\lambda$).
The metalizations have been assumed to consist of copper with conductivity
$\sigma=5.8\times10^{7}$ {[}S/m{]} and thickness $30\times10^{-6}$
{[}m{]}. The meta-atom lattice periodicity has been set to $\Delta=3.7$
{[}mm{]} (i.e., $\Delta\approx0.32\lambda$).%
\begin{figure}
\begin{center}\begin{tabular}{c}
\includegraphics[%
  width=0.87\columnwidth,
  keepaspectratio]{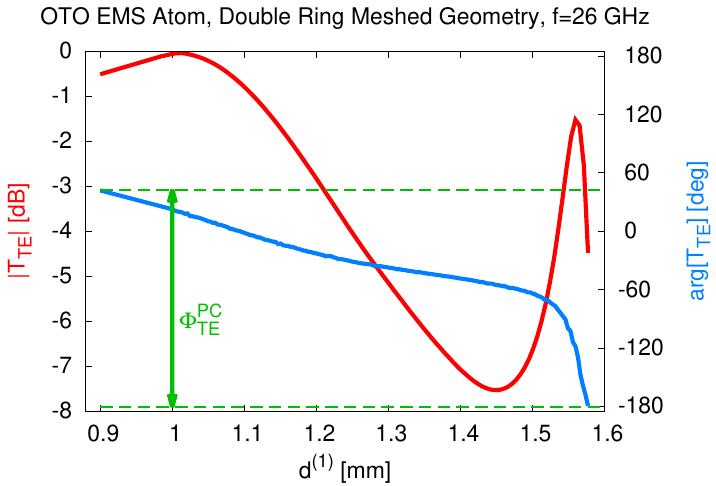}\tabularnewline
(\emph{a})\tabularnewline
\includegraphics[%
  width=0.87\columnwidth,
  keepaspectratio]{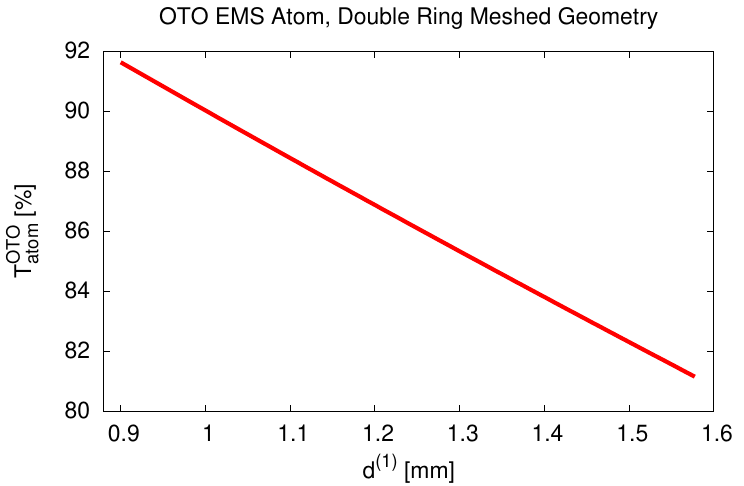}\tabularnewline
(\emph{b})\tabularnewline
\tabularnewline
\tabularnewline
\end{tabular}\end{center}

\caption{\emph{OTO Meta-Atom Design} ($f_{0}=26$ {[}GHz{]}) - Plot of (\emph{a})
the magnitude and the phase of $T_{TE}\left(\underline{d}\right)$
and (\emph{b}) the behaviour of the meta-atom transparency $\mathcal{T}_{atom}^{OTO}$
versus $d^{\left(1\right)}$. }
\end{figure}

\subsection{\noindent \emph{OTO} Meta-Atom Design\label{sub:OTO-Atom-Synthesis}}

\noindent Following the guidelines detailed in Sect. II.A, the first
task has been the design of an \emph{OTO} meta-atom enabling advanced
transmitted-field control properties subject to the constraint of
being based on a minimum-complexity structure featuring transparent
conducting layers patterned on a non-dedicated glass-based substrate.

\noindent Referring to the reference parametric model of the \emph{EMS}
unit cell in Fig. 2(\emph{a}), the setup of the meshed copper (i.e.,
the patterning conductor) has been chosen according to \cite{Kang 2018}
for yielding a mesh fill factor of about $\mathcal{F}_{mesh}\approx11.7\%$,
which corresponds to a single-layer conductor optical transmittance
equal to $\mathcal{T}_{mesh}=\left(1-\mathcal{F}_{mesh}\right)^{2}\approx77.8\%$.
More in detail, $d^{\left(3\right)}=30\times10^{-6}$ {[}m{]} (i.e.,
the thickness of the mesh wires is roughly half of a typical human
hair), $d^{\left(4\right)}=225\times10^{-6}$ {[}m{]}, and $d^{\left(5\right)}=20$
{[}deg{]}. The meta-atom model has been then parametrically tuned
to operate at $f_{0}$ by properly optimizing $d^{\left(1\right)}$
and $d^{\left(2\right)}$ {[}Fig. 2(\emph{a}){]}. By setting the trade-off
ring width value $d^{\left(2\right)}=795\times10^{-6}$ {[}m{]}, $d^{\left(1\right)}$
has been chosen as the control parameter for the entries of $\overline{\overline{T}}$
{[}i.e., $\overline{\overline{T}}\left(\underline{d}\right)=\overline{\overline{T}}\left(d^{\left(1\right)}\right)${]}.%
\begin{table}

\caption{\emph{OTO Meta-Atom Design} ($f_{0}=26$ {[}GHz{]}) - Characteristics
of the \emph{OTO} meta-atom and comparison with the state-of-the-art.}

\begin{center}{\footnotesize }\begin{tabular}{|c||c|c|c|c|c|c|}
\hline 
&
{\footnotesize $f_{0}$ }&
&
&
{\footnotesize $\mathcal{T}_{atom}^{OTO}$}&
{\footnotesize $\Phi_{TE}^{PC}$ }&
{\footnotesize $\Phi_{TE}^{MAG}$ }\tabularnewline
{\footnotesize Design}&
{\footnotesize {[}GHz{]}}&
{\footnotesize $N$}&
\emph{\footnotesize IG?}&
{\footnotesize {[}$\%${]} }&
{\footnotesize {[}deg{]}}&
{\footnotesize {[}dB{]}}\tabularnewline
\hline
\hline 
{\footnotesize Ideal $N=2$}&
{\footnotesize n.a.}&
{\footnotesize $2$}&
{\footnotesize n.a.}&
{\footnotesize n.a.}&
{\footnotesize $230$}&
{\footnotesize $-3.0$}\tabularnewline
\hline 
{\footnotesize \cite{Hong 2022}}&
{\footnotesize $38.5$}&
{\footnotesize $2$}&
{\footnotesize No}&
{\footnotesize $90$}&
{\footnotesize $123$}&
{\footnotesize $-1.5$}\tabularnewline
\hline 
{\footnotesize \cite{Kim 2023}}&
{\footnotesize $28$}&
{\footnotesize $2$}&
{\footnotesize No}&
{\footnotesize $95$}&
{\footnotesize $196$}&
{\footnotesize $-3.0$}\tabularnewline
\hline 
{\footnotesize \cite{Liu 2019}}&
{\footnotesize $28.5$}&
{\footnotesize $3$}&
{\footnotesize No}&
{\footnotesize n.a.}&
{\footnotesize $290$}&
{\footnotesize $-3.0$}\tabularnewline
\hline 
{\footnotesize This work}&
{\footnotesize $26$}&
{\footnotesize $2$}&
{\footnotesize Yes}&
{\footnotesize $80$}&
{\footnotesize $220$}&
{\footnotesize $-7.7$}\tabularnewline
\hline
\end{tabular}\end{center}
\end{table}

\noindent Figure 4(\emph{a}) summarizes the transmission performance
of the optimized meta-atom structure. It is worthwhile to point out
that they have been predicted without considering equivalent homogenized
models for the wire mesh (i.e., the actual copper wire structure has
been simulated in \emph{Ansys HFSS}). It turns out that when varying
$d^{\left(1\right)}$ within its feasibility range $d_{\min}^{\left(1\right)}\leq d^{\left(1\right)}\leq d_{\max}^{\left(1\right)}$
being $d_{\min}^{\left(1\right)}=0.9\times10^{-3}$ {[}m{]} and $d_{\max}^{\left(1\right)}=1.6\times10^{-3}$,
such a unit cell supports a phase coverage of approximately $\Phi_{TE}^{PC}\approx220$
{[}deg{]} with a worst-case transmittance magnitude of $\Phi_{TE}^{MAG}\approx-7.7$
{[}dB{]} {[}Fig. 4(\emph{a}){]}%
\footnote{\noindent The same behavior holds true for the \emph{TM} component
owing to the cell symmetry, while the magnitude of the cross-polar
components turns out $<-30$ {[}dB{]}.%
}. The effects of the meta-atom insertion loss on the \emph{OTO-EMS}s
performance are assessed in the subsequent sub-section. However, let
us notice that the theoretical maximum phase range enabled by a non-transparent
ideal two-layer meta-atom is around $230$ {[}deg{]} for a $50\%$
transmission efficiency \cite{Yang 2019}, \cite{Abdelrahman 2014}
(Tab. I). Potential solutions to improve such performance while keeping
the glass panel configuration may thus require increasing the number
meta-atom layers as well as employing optimized inter-layer spacings
\cite{Abdelrahman 2014}.%
\begin{figure}
\begin{center}\begin{tabular}{cc}
\includegraphics[%
  width=0.45\columnwidth,
  keepaspectratio]{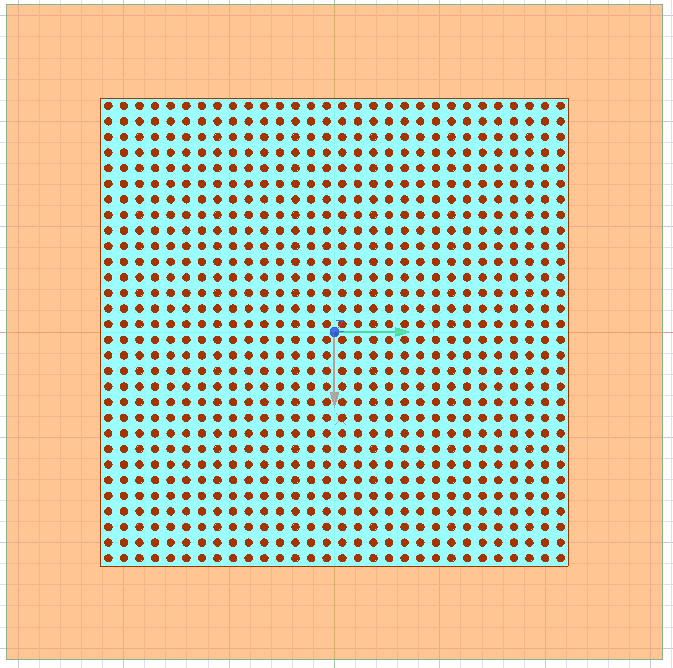}&
\includegraphics[%
  width=0.45\columnwidth,
  keepaspectratio]{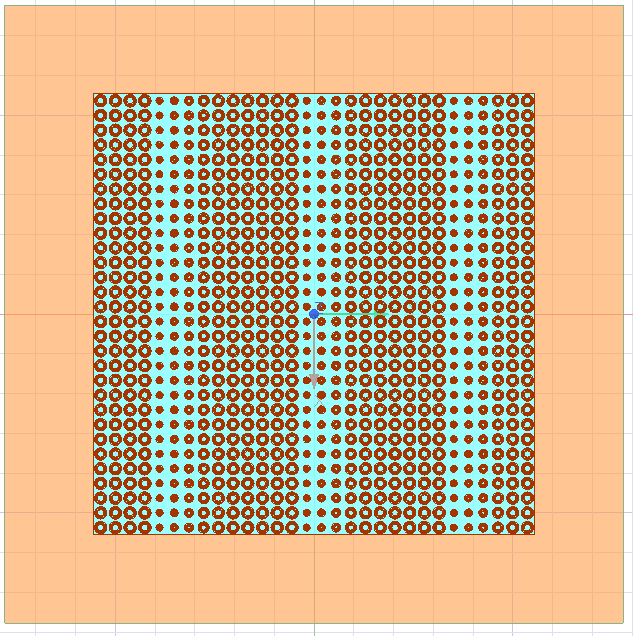}\tabularnewline
(\emph{a})&
(\emph{b})\tabularnewline
\includegraphics[%
  width=0.45\columnwidth,
  keepaspectratio]{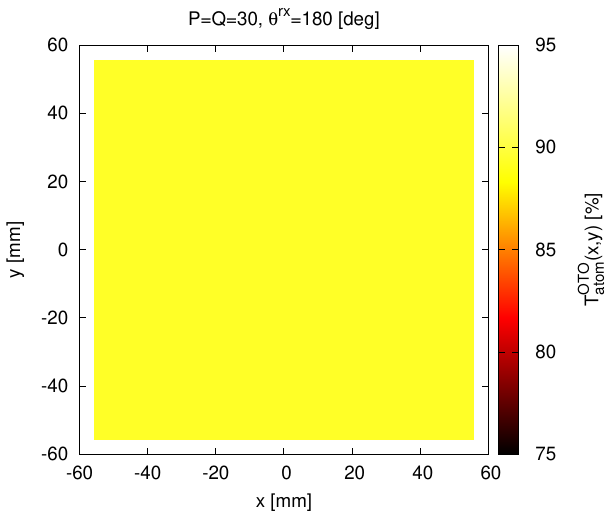}&
\includegraphics[%
  width=0.45\columnwidth,
  keepaspectratio]{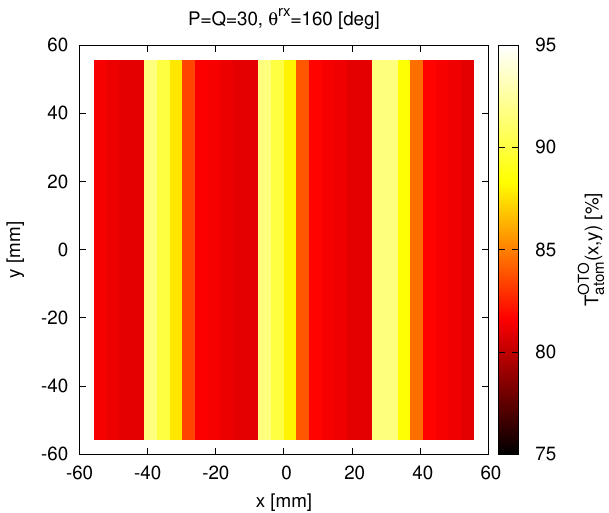}\tabularnewline
(\emph{c})&
(\emph{d})\tabularnewline
\end{tabular}\end{center}

\caption{\emph{OTO-EMS Layout Synthesis} ($f_{0}=26$ {[}GHz{]}, $P=Q=30$)
- Plot of (\emph{a})(\emph{b}) the \emph{OTO-EMS} layouts and (\emph{c})(\emph{d})
the corresponding optical transparency index when (\emph{a})(\emph{c})
$\theta^{rx}=180$ {[}deg{]} and (\emph{b})(\emph{d}) $\theta^{rx}=160$
{[}deg{]}.}
\end{figure}

\noindent As regards the relation between $\overline{\overline{T}}$
and $d^{\left(1\right)}$ {[}Fig. 4(\emph{a}){]}, it is worth noticing
that the observed sensitivity is consistent with that of standard
meta-atom architectures at such frequencies \cite{Oliveri 2023F},
\cite{Hong 2022}, \cite{Liu 2019} and compliant with current generation
\emph{PCB} fabrication processes.

\noindent As for the transparency of the arising meshed meta-atom,
the dependence of its optical transmittance on $d^{\left(1\right)}$
is shown in Fig. 4(\emph{b}). The plot indicates that $\mathcal{T}_{atom}^{OTO}>80\%$
regardless of the meta-atom setup within its feasibility range (i.e.,
$\underline{d}\in\Psi^{opt}$). This confirms the potentialities of
meshed copper meta-atoms \cite{Sharma 2019}-\cite{Silva 2021b} for
building printed optically-transparent \emph{EMS}s {[}Fig. 4(\emph{b}){]}.
A comparison of the essential performance metrics and compliancy with
\emph{IG} installation (i.e., {}``\emph{IG}?'' column in Tab. I)
of the proposed \emph{OTO} meta-atom with those of state-of-the-art
transparent architectures featuring different number of metalization
layers $N$ is provided in Tab. I, for the sake of completeness.

\subsection{\noindent \emph{OTO-EMS} Layout Synthesis\label{sub:OTO-EMS-Design}}

\noindent The first test case of this section, which is devoted to
analyze the \emph{O2I} transmission performance of \emph{OTO-EMS}s,
deals with the synthesis of a square \emph{EMS} $\mathcal{L}\approx11.1$
{[}cm{]}-sided ($P=Q=30$) transmitting the wave towards $\left(\theta^{rx},\varphi^{rx}\right)=\left(180,0\right)$
{[}deg{]} (i.e., no anomalous transmission). The \emph{OTO-EMS} layout,
synthesized with the procedure of Sub-Sect. II.B, presents a uniform
pattern with all cells identical {[}Fig. 5(\emph{a}){]} as expected
from the Generalized Snell's Law for refraction \cite{Yang 2019}.
The corresponding local optical transparency map {[}Fig. 5(\emph{c}){]}
is uniform, as well, with a constant value of $\mathcal{T}_{atom}^{OTO}$
around $90\%$ across the whole \emph{EMS} support $\Omega$. %
\begin{figure}
\begin{center}\includegraphics[%
  width=0.95\columnwidth,
  keepaspectratio]{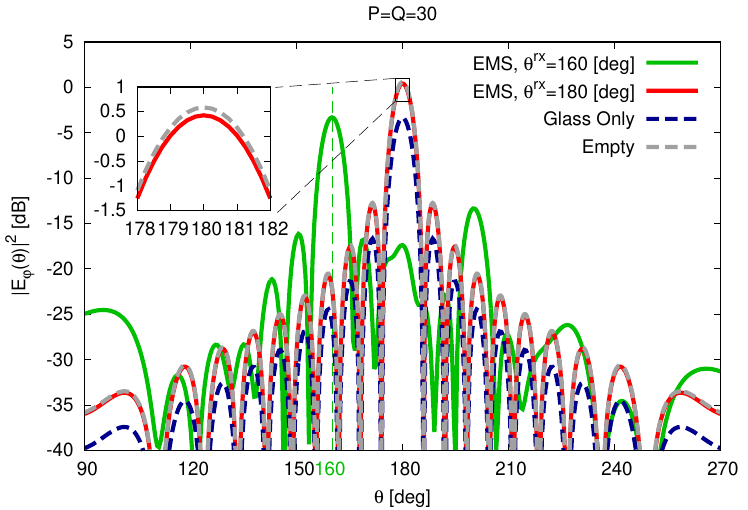}\end{center}

\caption{\emph{OTO-EMS Layout Synthesis} - ($f_{0}=26$ {[}GHz{]}, $P=Q=30$,
$\varphi^{rx}=0$ {[}deg{]}) - Plots of the transmitted pattern, $\left|E_{\varphi}^{tran}\left(\mathbf{r}\right)\right|^{2}$,
versus $\theta$.}
\end{figure}

\noindent Figure 6 compares the behaviour of the magnitude%
\footnote{\noindent In the following, no normalization has been applied to the
plotted quantities (i.e., the far-region electrical field magnitude
has been reported) unless otherwise specified.%
} of the dominant $\varphi$-component of the electric field $\left|E_{\varphi}^{tran}\left(\mathbf{r}\right)\right|^{2}$
transmitted in the horizontal cut {[}$\varphi=0$ {[}deg{]} - Fig.
1(\emph{a}){]} when applying the \emph{OTO-EMS} in Fig. 5(\emph{a})
{[}Fig. 6 (red line){]} and when considering either non-patterned
\emph{IG} panel of the same size {[}{}``Glass only'' - Fig. 6 (blue
dashed line){]} or when considering a hollow region of identical size
$\Omega$ {[}{}``Empty'' - Fig. 6 (grey dashed line){]}. One can
clearly infer that the \emph{OTO-EMS} patterning considerably improves
the transmitted power with respect to the standard commercial \emph{IG}
window of the same size (i.e., $\frac{\left|E_{\varphi}^{EMS}\left(\theta\right)\right|_{\theta=180\, deg}^{2}}{\left|E_{\varphi}^{Glass}\left(\theta\right)\right|_{\theta=180\, deg}^{2}}\approx3.7$
{[}dB{]} - Fig. 6). Moreover, interested readers should notice that,
despite the architectural and material constraints, such a meshed
\emph{EMS} features a power loss only $0.16$ {[}dB{]} below the ideal
hollow case (Fig. 6), while the beam-width size is equal since it
depends on the panel aperture $\Omega$.%
\begin{figure}
\begin{center}\begin{tabular}{c}
\includegraphics[%
  width=0.90\columnwidth,
  keepaspectratio]{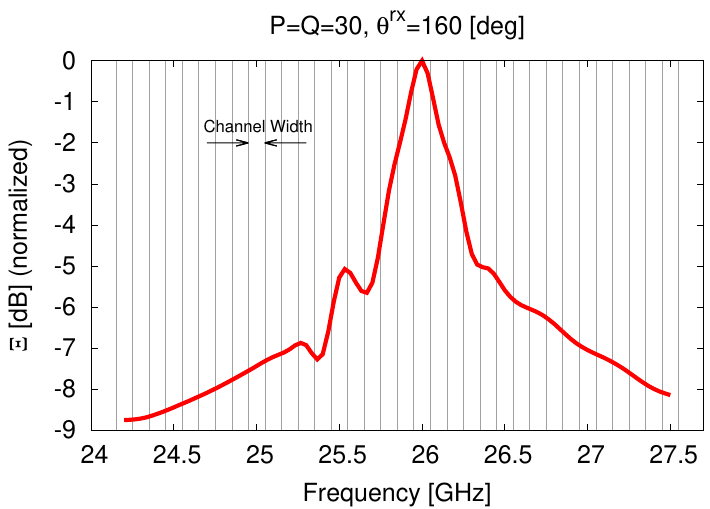}\tabularnewline
(\emph{a})\tabularnewline
\includegraphics[%
  width=0.90\columnwidth,
  keepaspectratio]{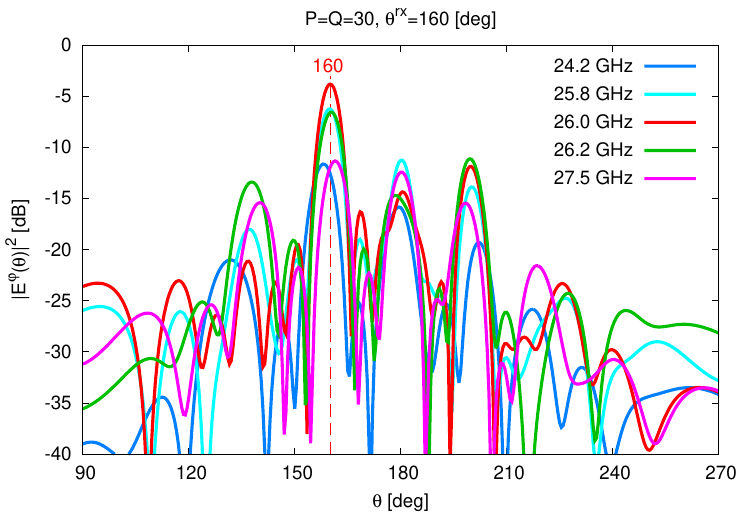}\tabularnewline
(\emph{b})\tabularnewline
\end{tabular}\end{center}

\caption{\emph{OTO-EMS Layout Synthesis} - ($f_{0}\in\left[24.2,27.5\right]$
{[}GHz{]}, $P=Q=30$, $\theta^{rx}=160$ {[}deg{]}, $\varphi^{rx}=0$
{[}deg{]}) - Plots of (\emph{a}) the peak power pattern, $\Xi$, versus
$f_{0}$ and (\emph{b}) the transmitted pattern, $\left|E_{\varphi}^{tran}\left(\mathbf{r}\right)\right|^{2}$,
versus $\theta$.}
\end{figure}

\noindent The possibility to derive an \emph{OTO-EMS} that supports
anomalous transmission angles, with an adequate control of the transmitted
beam, is validated next by keeping the same scenario of the previous
example, but now considering an anomalous transmission angle of $\Delta\theta^{rx}=20$
{[}deg{]} ($\Delta\theta^{rx}\triangleq180-\theta^{rx}$; $\theta^{rx}=160$
{[}deg{]}). Figure 5(\emph{b}) shows the synthesized \emph{OTO-EMS}
layout, while the plot of $\mathcal{T}_{atom}^{OTO}\left(x,y\right)$
is reported in Fig. 5(\emph{d}). This latter indicates that an excellent
local transparency is achieved across the entire \emph{EMS} support
$\Omega$ since always $\mathcal{T}_{atom}^{OTO}>80\%$. More importantly,
the transmitted pattern along the horizontal cut in Fig. 6 (green
line) proves that the \emph{OTO-EMS} focuses the energy towards the
anomalous direction by yielding an acceptable scan loss ($\frac{\left|E_{\varphi}^{EMS}\left(\theta\right)\right|_{\theta=160[deg]}^{2}}{\left|E_{\varphi}^{EMS}\left(\theta\right)\right|_{\theta=180[deg]}^{2}}\approx-3.6$
{[}dB{]} - Fig. 6), which is unavoidable because of the planar nature
of the \emph{EMS} as well as the insertion loss performance of the
designed meta-atom {[}Fig. 4(\emph{a}){]}. 

\noindent The presence of moderate quantization side-lobes (e.g.,
$\theta\approx200$ {[}deg{]} - Fig. 6), unlike the {}``non-anomalous
transmission'' case, is due to the limited phase coverage of the
two-layer meta-atom at hand {[}Fig. 4(\emph{a}){]}. In fact, the designed
\emph{OTO-EMS}, featuring $14$ different unit cell configurations%
\footnote{\noindent The number of different unit cells has not been constrained
in the synthesis process, unlike the methods in \cite{Hong 2022},
\cite{Kim 2023}, but likewise many state-of-the-art design strategies
for static passive \emph{EMS}s \cite{Yang 2019}, \cite{Oliveri 2021c}-\cite{Oliveri 2023b},
\cite{Oliveri 2023}, \cite{Mencagli 2020}, \cite{Oliveri 2023F},
\cite{Liu 2019}, \cite{Huang 2008}.%
}, yields an average $12.7$ {[}deg{]} phase error with respect to
the optimal phase profile. Of course, a better side-lobe control may
be obtained whether exploiting multi-layer structures, but this would
not be compliant with the retro-fitting constraint of the problem
at hand.%
\begin{figure}
\begin{center}\includegraphics[%
  width=0.95\columnwidth,
  keepaspectratio]{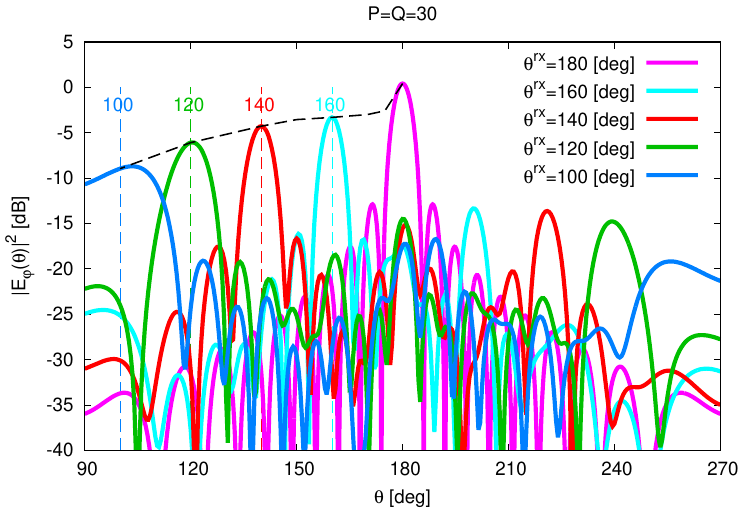}\end{center}

\caption{\emph{OTO-EMS Layout Synthesis} - ($f_{0}=26$ {[}GHz{]}, $P=Q=30$,
$\varphi^{rx}=0$ {[}deg{]}) - Plots of the transmitted pattern, $\left|E_{\varphi}^{tran}\left(\mathbf{r}\right)\right|^{2}$,
versus $\theta$.}
\end{figure}

\noindent The bandwidth performance of the proposed \emph{OTO-EMS}s
architecture is assessed in the subsequent numerical experiment. To
this end, the previous $\Delta\theta^{rx}=20$ {[}deg{]} design is
analyzed considering the n258 5G \emph{mmW} band (i.e., $f_{0}\in\left[24.2,27.5\right]$
{[}GHz{]}). The plot of the peak power pattern $\Xi$ (i.e., $\Xi\triangleq\max_{\theta,\varphi}\left|E_{\varphi}^{tran}\left(\mathbf{r}\right)\right|^{2}$)
shows that, as expected owing to the resonant nature of the conceived
meta-atom (Fig. 2), the skin efficiency reduces when the the illumination
frequency deviates from the nominal $f=26$ {[}GHz{]}, with a worst-case
transmitted power reduction of $\approx9$ {[}dB{]} at the n258 band
edges {[}Fig. 7(\emph{a}){]}. However, such losses are below $3$
dB in $5$ n258 band channels {[}i.e., $100$ MHz-wide bands centered
in the carrier frequencies $f_{0}=\left\{ 25.8,25.9,26.0,26.1,26.2\right\} $
{[}GHz{]} - Fig. 7(\emph{a}){]}. It is worth remarking that such results
have been obtained without re-optimizing the unit cell architecture
(Fig. 2) on a broader band. Moreover, the designed structure still
exhibits a collimated transmission (i.e., a well-defined main lobe)
in the desired $\theta^{rx}$ direction in the entire n258 band {[}e.g.,
$f_{0}=24.2$ {[}GHz{]}; $f_{0}=27.5$ {[}GHz{]} - Fig. 7(\emph{b}){]}.

\noindent The next test case analyzes the performance degradation
versus the scan angle. Towards this end, a set of \emph{OTO-EMS}s
has been designed by varying the anomalous transmission angle in the
range $\Delta\theta^{rx}\in\left\{ 40,\,60,\,80\right\} $ {[}deg{]}
($\theta^{rx}\in\left\{ 140,\,120,\,100\right\} $ {[}deg{]}). By
comparing the transmitted beams in Fig.  8, the following outcomes
can be drawn: (\emph{i}) it is possible to yield an effective beam
focusing despite the limited phase coverage of the \emph{OTO} meta-atom
and even when the scan angle is close to end-fire {[}e.g., $\Delta\theta=80$
{[}deg{]}, $\theta^{rx}=100$ {[}deg{]} - Fig.  8 (blue line){]};
(\emph{ii}) regardless of the scan angle, the presence of moderate
quantization lobes is confirmed, the major side-lobe usually appearing
specularly to the orthogonal transmission angle (i.e., $\theta\approx180+\Delta\theta^{rx}$);
(\emph{iii}) as expected, there is the unavoidable beam broadening
effect when the transmission angle gets closer to end-fire (e.g.,
blue vs. red line - Fig.  8) as well as an increase of the scan loss.
\begin{figure}
\begin{center}\includegraphics[%
  width=0.95\columnwidth,
  keepaspectratio]{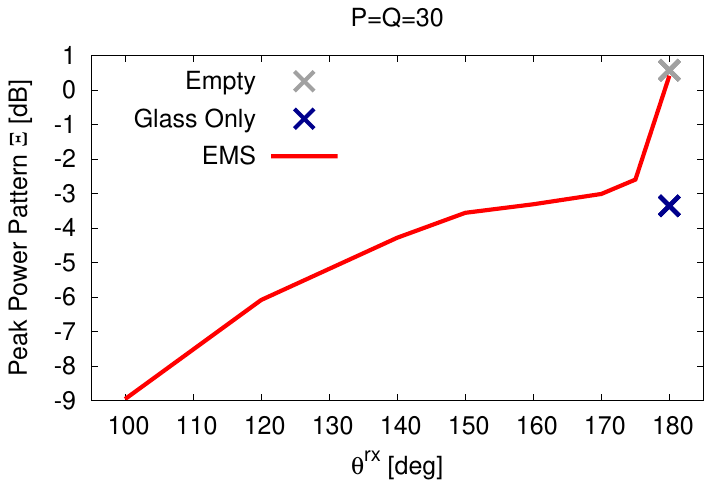}\end{center}

\caption{\emph{OTO-EMS Layout Synthesis} - ($f_{0}=26$ {[}GHz{]}, $P=Q=30$,
$\varphi^{rx}=0$ {[}deg{]}) - Plots of the peak power pattern, $\Xi$,
versus $\theta^{rx}$.}
\end{figure}
This latter phenomenon is pointed out in Fig.  9 where the plot of
the peak power pattern $\Xi$ versus $\theta^{rx}$ is shown. As it
can be observed, the scan loss sharply grows as $\theta^{rx}$ deviates
from $\theta^{rx}=180$ {[}deg{]}. This is actually caused by the
presence of the quantization lobes in the beam pattern as well as
the lower efficiency caused by insertion loss of the different unit
cells {[}Fig. 4(\emph{a}){]}, unlike the {}``non-anomalous transmission''
case where all the meta-atoms are identical and therefore the design
can be implemented exploiting the meta-atom with the best wireless
efficiency {[}e.g., Fig. 8 (magenta line){]}.

\noindent The transmitted pattern of a window only partially covered
by an \emph{OTO-EMS} is numerically evaluated next. Towards this end,
the previously designed $P=Q=30$, $\Delta\theta^{rx}=60$ {[}deg{]}
($\theta^{rx}=120$ {[}deg{]}) skin is assumed to be installed at
the center of a window pane of aperture $0.6\times1.1$ {[}$\mathrm{m}^{2}${]}.
The plot of the transmitted patterns through the window alone (green
line - Fig. 9), the \emph{OTO-EMS} alone (red line - Fig. 10), or
through their combination (red line - Fig. 10) shows that even a small
skin panel (i.e., $\mathcal{L}\approx11.1$ {[}cm{]}) enables to generate
a collimated beam in the desired $\theta^{rx}$ direction which is
$>37$ dB above the power pattern generated by the window itself in
$\theta=\theta^{rx}$ and approximately $8$ dB above the sidelobe
envelope (blue vs. green line - Fig. 10).%
\begin{figure}
\begin{center}\includegraphics[%
  width=0.90\columnwidth,
  keepaspectratio]{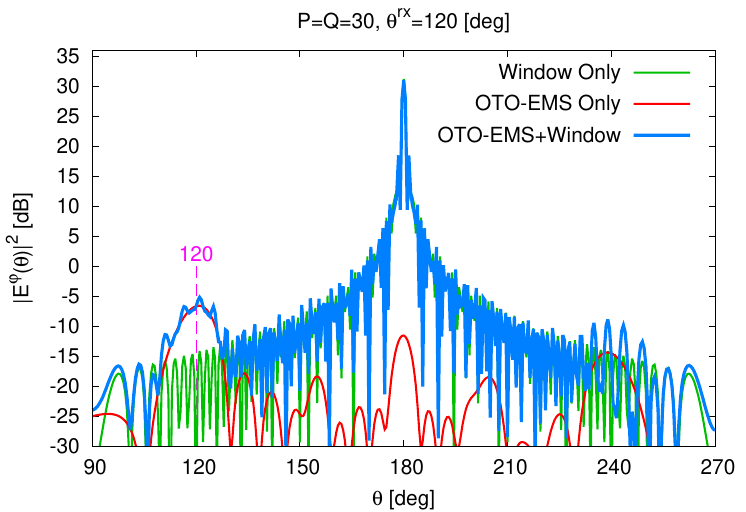}\end{center}

\caption{\emph{OTO-EMS Layout Synthesis} - ($f_{0}=26$ {[}GHz{]}, $P=Q=30$,
$\theta^{rx}=120$ {[}deg{]}, $\varphi^{rx}=0$ {[}deg{]}) - Plots
of the transmitted pattern, $\left|E_{\varphi}^{tran}\left(\mathbf{r}\right)\right|^{2}$,
versus $\theta$ when considering a glass window pane of aperture
$0.6\times1.1$ {[}$\mathrm{m}^{2}${]}.}
\end{figure}

\noindent The behaviour of the conceived \emph{OTO-EMS}s when dealing
with oblique incidence angles is addressed in the subsequent numerical
experiment. To this end, the design process has been carried out assuming
$\theta^{inc}=-40$ {[}deg{]} and setting $\Delta\theta^{rx}\in\left\{ 0,\,20,\,40,\,60,\,80\right\} $
{[}deg{]} ($\theta^{rx}\in\left\{ 180,\,160,\,140,\,120,\,100\right\} $
{[}deg{]}). The results reported in Fig. 11 indicates that, despite
the unavoidable quantization lobe in the $\theta=140$ {[}deg{]} direction
caused by the phase coverage limitations of the two-layer meta-atom
at hand {[}Fig. 4(\emph{a}){]}, the synthesized \emph{OTO-EMS}s always
support a collimated transmitted beam in the $\theta^{rx}$ direction
even though $\theta^{inc}\neq0$ {[}deg{]} (Fig. 11).%
\begin{figure}
\begin{center}\includegraphics[%
  width=0.90\columnwidth,
  keepaspectratio]{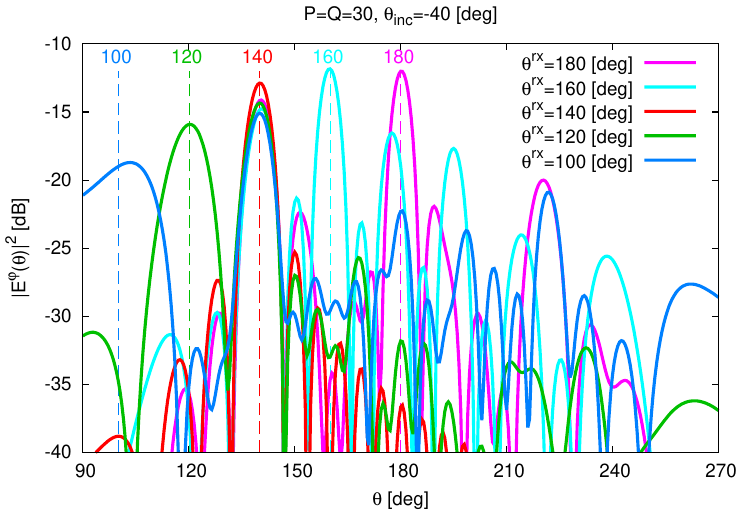}\end{center}

\caption{\emph{OTO-EMS Layout Synthesis} - ($f_{0}=26$ {[}GHz{]}, $P=Q=30$,
$\left(\theta^{inc},\varphi^{inc}\right)=\left(-40,0\right)$ {[}deg{]},
$\varphi^{rx}=0$ {[}deg{]}) - Plots of the transmitted pattern, $\left|E_{\varphi}^{tran}\left(\mathbf{r}\right)\right|^{2}$,
versus $\theta$.}
\end{figure}

\noindent In the next example, the dependence of the transmission/focusing
performance on the panel aperture is assessed. By setting $\Delta\theta^{rx}=40$
{[}deg{]} (i.e., $\theta^{rx}=140$ {[}deg{]}), the panel side has
been varied in the range $7.4\leq\mathcal{L}\leq74$ {[}cm{]} (i.e.,
$20\leq P\leq200$ and $P=Q$). The plots of the peak power pattern
$\Xi$ versus the aperture size {[}Fig. 12(\emph{a}){]} indicate that
the transmitted power focused by the \emph{OTO-EMS} grows proportionally
to the panel aperture as logically expected {[}e.g., $\frac{\Xi_{\mathcal{L}=74\, cm}}{\Xi_{\mathcal{L}=7.4\, cm}}\approx39.9$
dB - Fig. 12(\emph{a}){]}. On the other hand, the comparison with
the maximum power transmitted along broadside by either a non-patterned
\emph{IG} panel of the same size {[}{}``Glass only'' - Fig. 12(\emph{a})
(blue line){]} or by a hollow region of identical aperture {[}{}``Empty''
- Fig. 12(\emph{a}) (grey line){]} provides a key proof of the effectiveness
of the \emph{OTO-EMS} concept in \emph{mmW} \emph{O2I} communications.
Indeed, it turns out that the synthesized patterned windows support
transmission along anomalous angles (e.g., $\Delta\theta^{rx}=40$
{[}deg{]}) with a power focusing efficiency which is less than $5$
{[}dB{]} below the theoretical optimum (i.e., the power transmitted
in the Snell's direction through the hollow region), despite the unavoidable
scan losses and the limited phase coverage of the meta-atom fulfilling
retrofitting constraints.%
\begin{figure}
\begin{center}\begin{tabular}{c}
\includegraphics[%
  width=0.95\columnwidth,
  keepaspectratio]{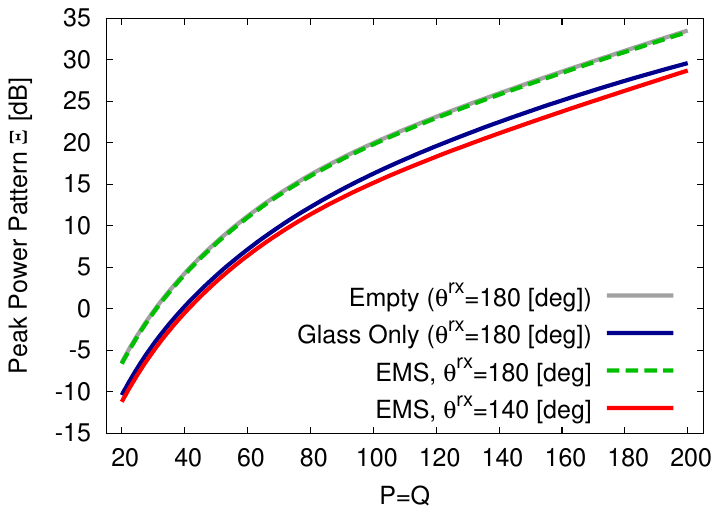}\tabularnewline
(\emph{a})\tabularnewline
\includegraphics[%
  width=0.95\columnwidth,
  keepaspectratio]{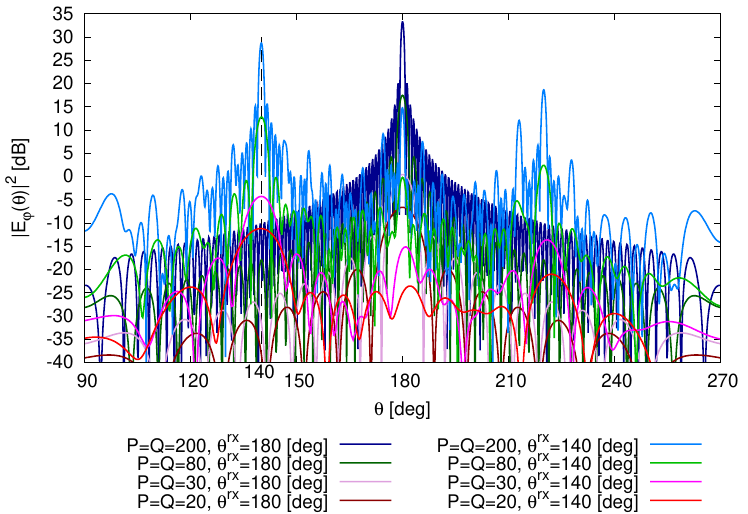}\tabularnewline
(\emph{b})\tabularnewline
\tabularnewline
\tabularnewline
\end{tabular}\end{center}

\caption{\emph{OTO-EMS Layout Synthesis} - ($f_{0}=26$ {[}GHz{]}, $\varphi^{rx}=0$
{[}deg{]}) - Behaviour of (\emph{a}) the peak power pattern, $\Xi$,
versus $P$ ($P=Q$) and plot of (\emph{b}) the transmitted pattern,
$\left|E_{\varphi}^{tran}\left(\mathbf{r}\right)\right|^{2}$, versus
$\theta$ when $\Delta\theta^{rx}\in\left\{ 0,40\right\} $ {[}deg{]}.}
\end{figure}

\noindent For completeness, the plots of the transmitted beam patterns
of representative apertures {[}i.e., $P\in\left\{ 20,30,80,200\right\} $
and $P=Q${]} are reported together with the broadside case (i.e.,
$\Delta\theta^{rx}=0$ {[}deg{]}) in Fig. 12(\emph{b}) to confirm
previous considerations on (\emph{a}) the presence and the location
of the quantization lobes, (\emph{b}) the beam-width dependence on
the panel aperture, and (\emph{c}) the improvement of the power transmission
when widening $\Omega$ {[}Fig. 12(\emph{b}){]}.%
\begin{figure}
\begin{center}\begin{tabular}{c}
\includegraphics[%
  width=0.90\columnwidth,
  keepaspectratio]{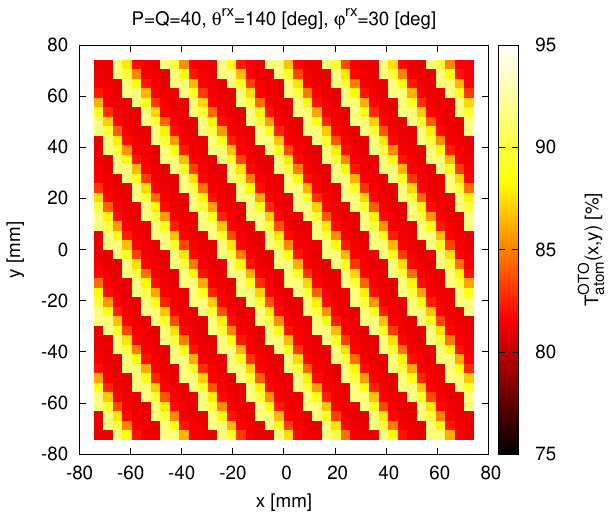}\tabularnewline
(\emph{a})\tabularnewline
\includegraphics[%
  width=0.90\columnwidth,
  keepaspectratio]{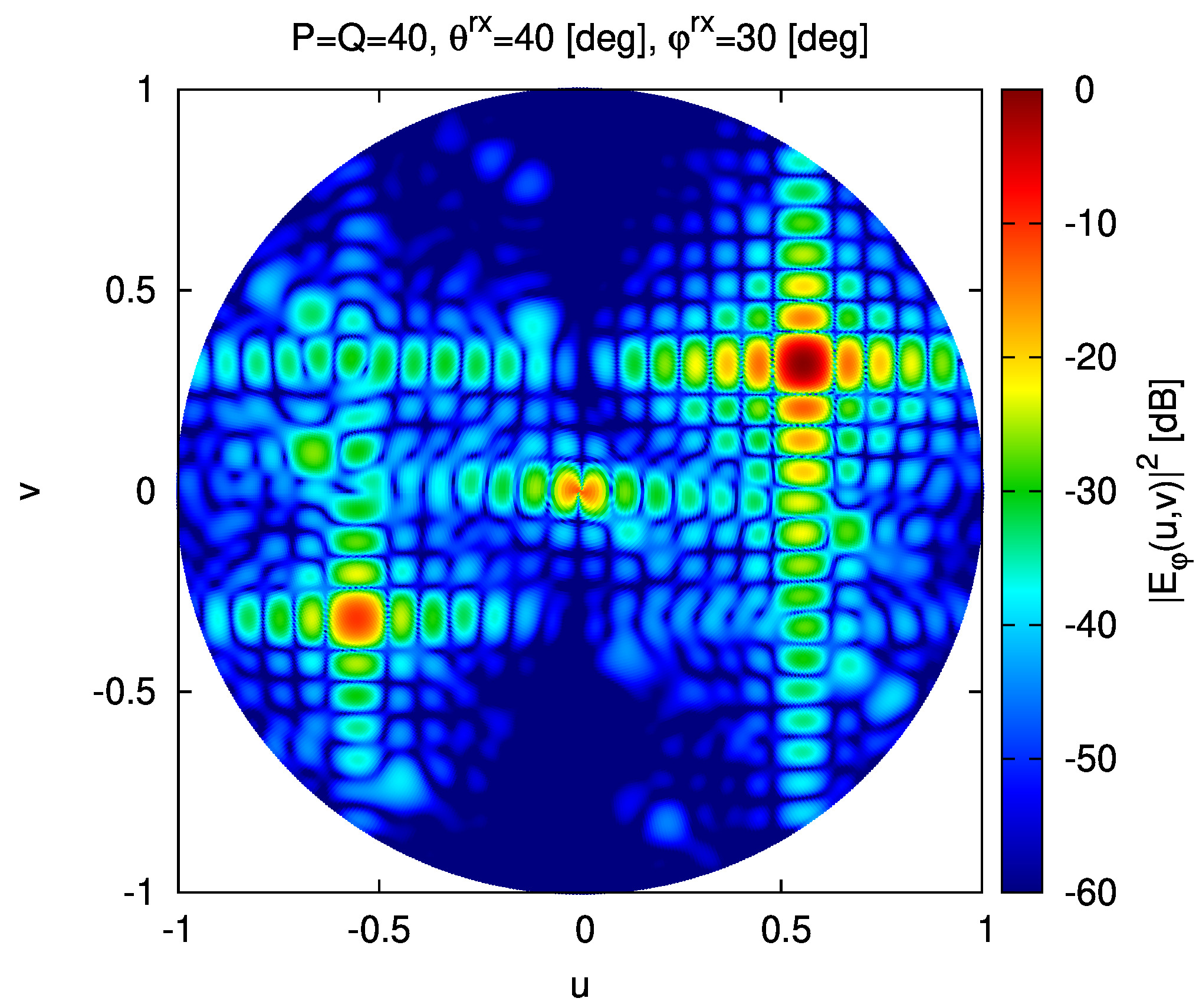}\tabularnewline
(\emph{b})\tabularnewline
\tabularnewline
\tabularnewline
\end{tabular}\end{center}

\caption{\emph{OTO-EMS Layout Synthesis} - ($f_{0}=26$ {[}GHz{]}, $P=Q=30$,
$\theta^{rx}\in140$ {[}deg{]}, $\varphi^{rx}=30$ {[}deg{]}) - Plot
of (\emph{a}) the profile of the optical transparency index within
$\Omega$ and (\emph{b}) the transmitted pattern, $\left|E_{\varphi}^{tran}\left(\mathbf{r}\right)\right|^{2}$,
in the $u-v$ plane.}
\end{figure}

\noindent The subsequent numerical experiment deals with a more complex
setup, the impinging wave being deflected in a double-anomalous direction
(i.e., a non-Snell direction both in elevation and azimuth). More
in detail, a $P=Q=40$ meta-atoms \emph{OTO-EMS} has been designed
by imposing the \emph{O2I} transmission along the direction ($\theta^{rx}=140$
{[}deg{]}, $\varphi^{rx}=30$ {[}deg{]}). Also in this case, the synthesized
\emph{OTO-EMS} features a local optical transparency greater than
$80\%$ within the entire aperture $\Omega$ {[}Fig. 13(\emph{a}){]}.
Moreover, the color map of the transmitted pattern in the $u-v$ coordinates
shows that the \emph{OTO-EMS} focuses the transmitted beam along the
desired direction since $u^{rx}\approx0.556$ and $v^{rx}\approx0.321$
{[}Fig. 13(\emph{b}){]}, being $u\triangleq\sin\left(\theta\right)\cos\left(\varphi\right)$
and $v\triangleq\sin\left(\theta\right)\sin\left(\varphi\right)$
the cosine directions \cite{Mailloux 2005}. Because of the limited
phase coverage of the dual-layer meta-atom and analogously to single-anomalous
transmissions, a specular quantization side-lobe appears at $u=-u^{rx}$,
$v=-v^{rx}$. Such quantization lobes may be reduced by considering
more complex meta-atom architectures enabling wider phase coverages.

\noindent The exploitation of the same theoretical framework when
dealing with more complex illuminations is then analyzed. To this
end, a $15$ {[}dB{]}-gain horn antenna {[}Fig. 14(\emph{a}){]} located
at $100$ {[}m{]} from the \emph{OTO-EMS} and $\left(\theta^{inc},\varphi^{inc}\right)=\left(-10,0\right)$
{[}deg{]} has been used as a benchmark radiator to model a primary
source (e.g., a base station), and the \emph{OTO-EMS} design has been
carried out enforcing $\Delta\theta^{rx}\in\left\{ 0,\,20,\,40,\,60,\,80\right\} $
{[}deg{]} ($\theta^{rx}\in\left\{ 180,\,160,\,140,\,120,\,100\right\} $
{[}deg{]}) {[}Fig. 14(\emph{b}){]}. The plots of the resulting patterns
confirm  that the beam shaping capabilities of the \emph{OTO-EMS}s
do not depend on the type of illumination {[}Fig. 14(\emph{b}){]}.
In fact, the desired anomalous transmission is achieved regardless
of the incident field {[}Fig. 14(\emph{b}){]}.%
\begin{figure}
\begin{center}\begin{tabular}{c}
\includegraphics[%
  width=0.95\columnwidth,
  keepaspectratio]{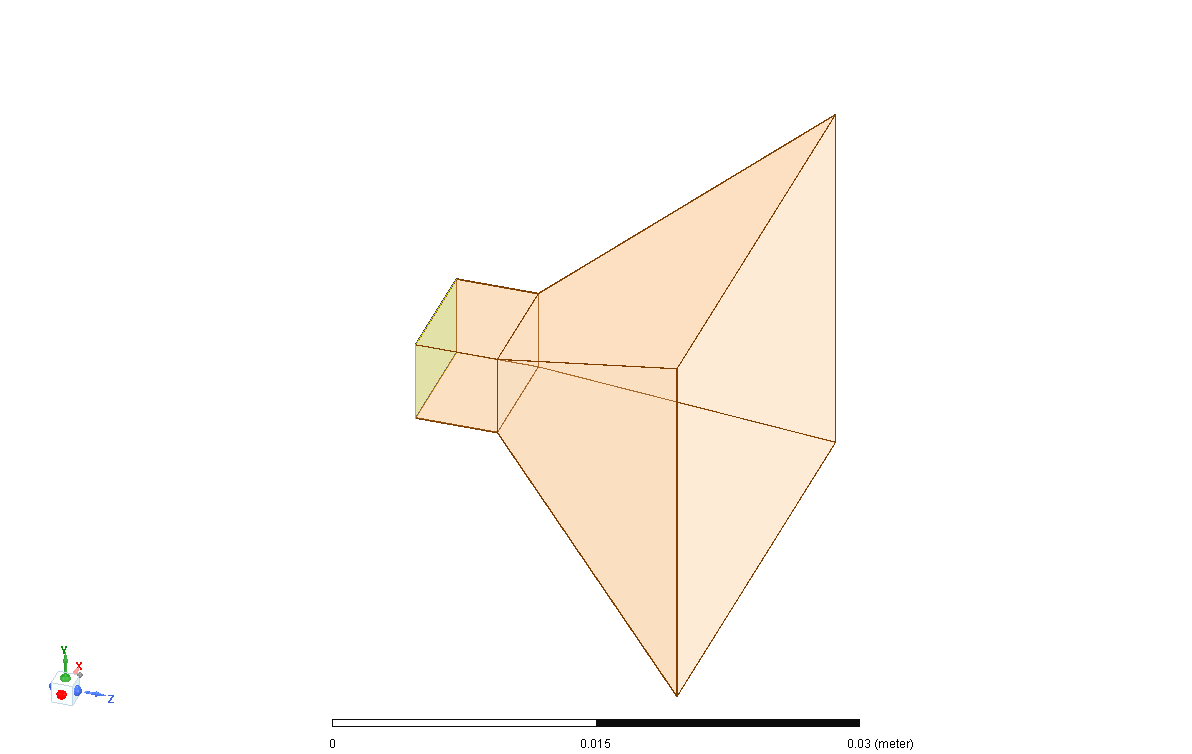}\tabularnewline
(\emph{a})\tabularnewline
\includegraphics[%
  width=0.90\columnwidth,
  keepaspectratio]{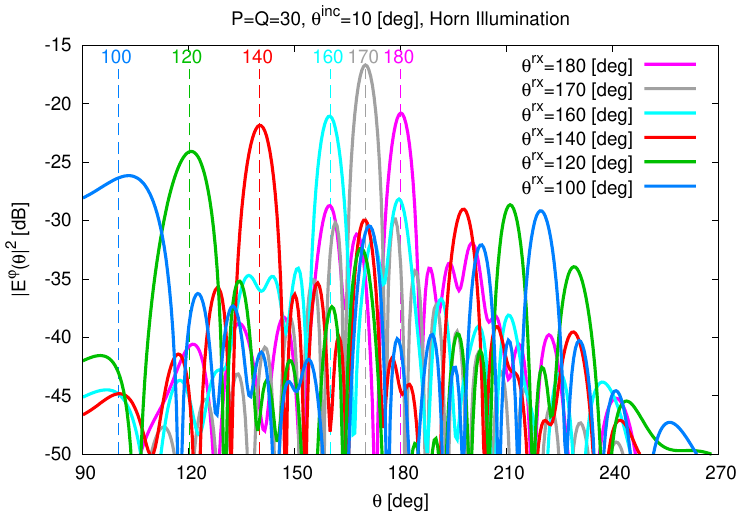}\tabularnewline
(\emph{b})\tabularnewline
\end{tabular}\end{center}

\caption{\emph{OTO-EMS Layout Synthesis} - ($f_{0}=26$ {[}GHz{]}, $P=Q=30$,
$\left(\theta^{inc},\varphi^{inc}\right)=\left(-10,0\right)$ {[}deg{]},
$\varphi^{rx}=0$ {[}deg{]}, \emph{Horn Illumination}) - Plots of
the transmitted pattern, $\left|E_{\varphi}^{tran}\left(\mathbf{r}\right)\right|^{2}$,
versus $\theta$.}
\end{figure}

\noindent The possibility to address more complex beamforming conditions
exploiting \emph{OTO-EMS}s has been investigated next. To this end,
a sidelobe level (\emph{SLL})-constrained \emph{OTO-EMS} synthesis
process following the guidelines in \cite{Oliveri 2021c} has been
implemented on a $P\times Q=30\times30$, $\left(\theta^{inc},\varphi^{inc}\right)=\left(0,0\right)$
{[}deg{]}, $\left(\theta^{rx},\varphi^{rx}\right)=\left(180,0\right)$
{[}deg{]} design. More in detail, the design has been carried out
enforcing a peak sidelobe $SLL\triangleq\frac{\max_{\theta\notin\Theta_{ML}}\left|E_{\varphi}^{tran}\left(\mathbf{r}\right)\right|^{2}}{\max_{\theta\in\Theta_{ML}}\left|E_{\varphi}^{tran}\left(\mathbf{r}\right)\right|^{2}}$
($\Theta_{ML}$ being the mainlobe region) of $SLL=-20$ dB\@. The
comparison of the normalized$\left|E_{\varphi}^{tran}\left(\mathbf{r}\right)\right|^{2}$
for the {}``unconstrained'' and {}``\emph{SLL} constrained'' designs
show that a \emph{SLL} mitigation can be achieved (i.e., $SLL_{constr}-SLL_{unconstr}=-5$
dB - Fig. 15) by slightly enlarging the mainlobe width, as theoretically
expected (Fig. 15) and already demonstrated for reflecting \emph{EMS}s
\cite{Oliveri 2021c}.%
\begin{figure}
\begin{center}\includegraphics[%
  width=0.90\columnwidth,
  keepaspectratio]{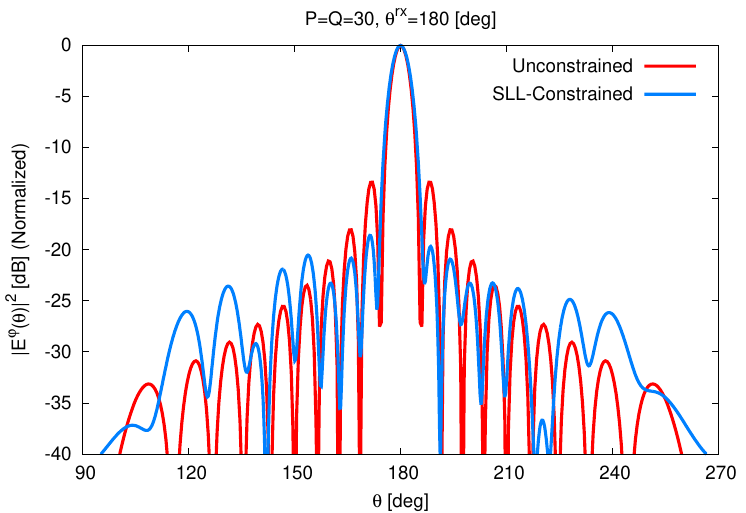}\end{center}

\caption{\emph{OTO-EMS Layout Synthesis} - ($f_{0}=26$ {[}GHz{]}, $P=Q=30$,
$\left(\theta^{rx},\varphi^{rx}\right)=\left(180,0\right)$ {[}deg{]})
- Comparison of the plots of the normalized transmitted pattern, $\left|E_{\varphi}^{tran}\left(\mathbf{r}\right)\right|^{2}$,
versus $\theta$, for unconstrained vs. sidelobe-constrained design.}
\end{figure}

\noindent The final test case is aimed at assessing the \emph{OTO-EMS}
performance in the presence of realistic non-ideal effects such as
diffraction from edges, surface waves, and non-periodic mutual coupling
that may arise in an \emph{OTO-EMS} prototype. Towards this end, a
full-wave analysis of an \emph{OTO-EMS} has been carried out exploiting
an industry-standard commercial simulator (\emph{Ansys HFSS} \cite{HFSS 2021}).
A finite \emph{OTO-EMS} model consisting of $P\times Q=20\times20$
cells {[}Fig. 16(\emph{a}){]} and designed to operate with $\left(\theta^{inc},\varphi^{inc}\right)=\left(-40,0\right)$
{[}deg{]}, $\left(\theta^{rx},\varphi^{rx}\right)=\left(160,0\right)$
{[}deg{]}) has been implemented considering a finite-element boundary-integral
(\emph{FE-BI}) formulation to avoid any numerical approximation resulting
from periodic boundary conditions. The plot of the analytically-computed
pattern positively compare with the full-wave simulated one {[}Fig.
16(\emph{b}){]}. More specifically, both the main beam location/magnitude
and the sidelobe location and envelope is confirmed by the full-wave
simulation {[}Fig. 16(\emph{b}){]}, with a minor mismatch in the angular
regions close to endfire possibly related to truncation effects and
surface waves {[}Fig. 16(\emph{b}){]}. Such a result, which is consistent
with the typical accuracy of state-of-the-art formulations \cite{Mencagli 2020},
supports the previous considerations regarding the \emph{OTO-EMS}
concept.%
\begin{figure}
\begin{center}\begin{tabular}{c}
\includegraphics[%
  width=0.95\columnwidth,
  keepaspectratio]{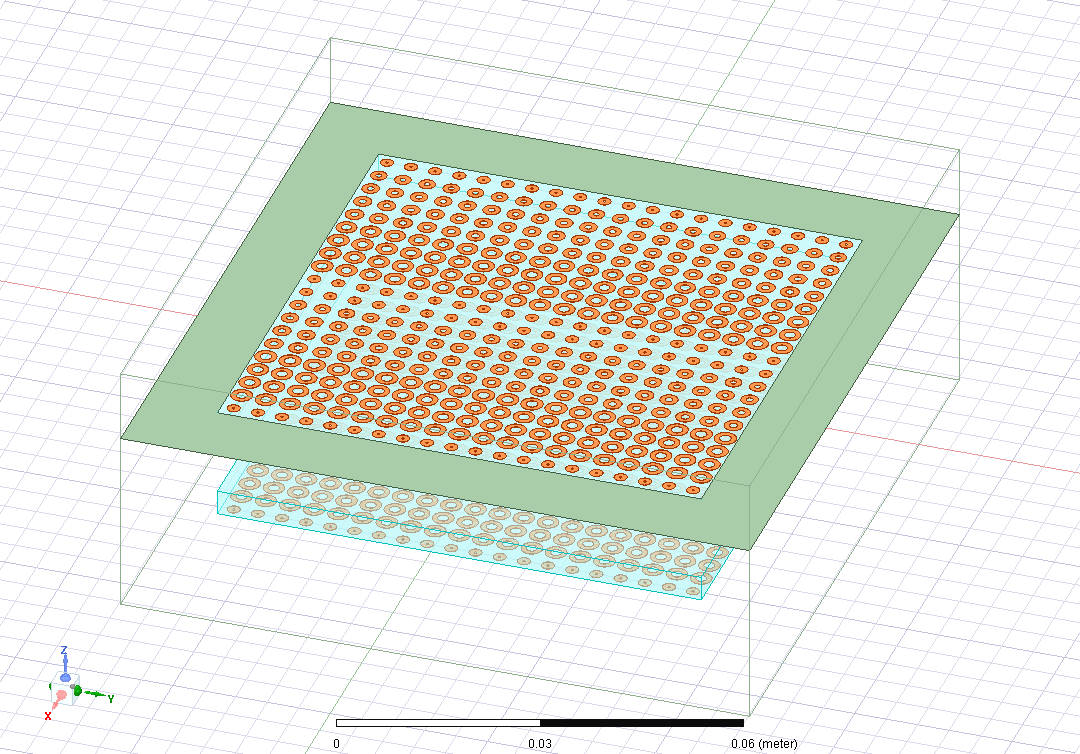}\tabularnewline
(\emph{a})\tabularnewline
\includegraphics[%
  width=0.90\columnwidth,
  keepaspectratio]{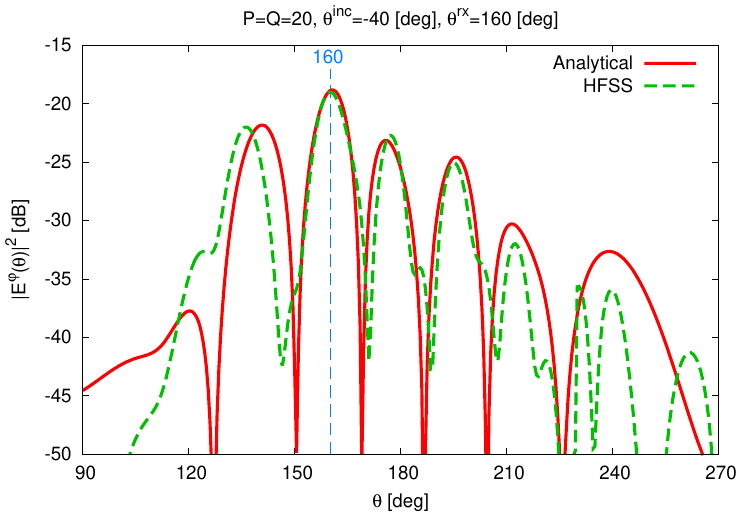}\tabularnewline
(\emph{b})\tabularnewline
\end{tabular}\end{center}

\caption{\emph{OTO-EMS Full-Wave Validation} - ($f_{0}=26$ {[}GHz{]}, $P=Q=20$,
$\left(\theta^{inc},\varphi^{inc}\right)=\left(-40,0\right)$ {[}deg{]},
$\left(\theta^{rx},\varphi^{rx}\right)=\left(160,0\right)$ {[}deg{]})
- 3D Model (\emph{a}) and plot of the analytical and Ansys HFSS-simulated
transmitted pattern, $\left|E_{\varphi}^{tran}\left(\mathbf{r}\right)\right|^{2}$,
versus $\theta$ (\emph{b}).}
\end{figure}

\section{\noindent Conclusions\label{sec:Conclusions-and-Remarks}}

\noindent \emph{OTO-EMS}s have been proposed as a profitable technological
solution to enable \emph{O2I} wireless communications at \emph{mmW}
frequencies. They consist of conducting optically-transparent patterned
layers attached to \emph{existing} glass windows to minimize both
costs and visual impacts by also allowing an easy deployment. Full-wave
numerical simulations of synthesized \emph{OTO-EMS}s of finite sizes
have assessed the feasibility as well as the effectiveness of \emph{O2I}
transmissions along both Snell and non-Snell refraction angles at
\emph{mmW}.

\noindent Accordingly, the fundamental methodological advancements
of this work include (\emph{i}) the demonstration of the feasibility
of mmWave transparent \emph{EMS}s suitable as a retro-fitting option
for existing windows / glass panels; (\emph{ii}) the assessment of
the \emph{OTO-EMS}s features in terms of wave control also in comparison
with non-patterned glass panels and addressing both standard and non-Snell
focusing as well as sidelobe mitigation; (\emph{iii}) the adaptation
of the \emph{SbD} synthesis concept to the case of transmitting \emph{EMS}
which generalizes existing implementations based on reflecting structures
\cite{Oliveri 2021c}-\cite{Oliveri 2023b}.

\noindent Future works, beyond the scope of the current paper and
also currently impossible in  our university owing to the lack of
suitable resources and facilities for the realization and the measurement
of the resulting devices at \emph{mmW} frequencies, will be aimed
at prototyping and measuring, in a controlled environment, the synthesized
\emph{OTO-EMS} samples. Towards this end, fabrication approaches previously
employed in metalens design such as \cite{Hong 2022}-\cite{Liu 2019}
may be generalized/customized to the scenario at hand. Nevertheless,
the presented full-wave simulation results obtained using an industry-standard
commercial software (\emph{Ansys HFSS} \cite{HFSS 2021}) provide
a preliminary evaluation of the impact of non-idealities such as diffraction
from edges, surface waves, and non-periodic mutual coupling on the
features of a practical \emph{OTO-EMS}.

\section*{\noindent Acknowledgements}

\noindent A. Massa wishes to thank E. Vico for her never-ending inspiration,
support, guidance, and help.

\begin{IEEEbiography}[{\includegraphics[width=1in,height=1.25in,clip,keepaspectratio]{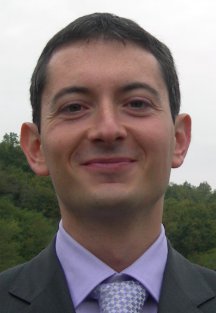}}]{Giacomo Oliveri} (IEEE Fellow) received the B.S. and M.S. degrees in Telecommunications Engineering and the PhD degree in Space Sciences and Engineering from the University of Genoa, Italy, in 2003, 2005, and 2009 respectively. He is currently an Associate Professor at the Department of Civil, Environmental, and Mechanical Engineering, University of Trento, and a Board Member of the ELEDIA Research Center. Moreover, he is Adjunct Professor at CentraleSupelec and member of the Laboratoire des signaux et systemes (L2S)@CentraleSupelec Gif-sur-Yvette (France). He has been a visiting researcher at L2S in 2012, 2013, and 2015, Invited Associate Professor at the University of Paris Sud, France, in 2014, and visiting professor at Universite Paris-Saclay in 2016 and 2017. He is author/co-author of over 400 peer reviewed papers on international journals and conferences. His research work is mainly focused on electromagnetic direct and inverse problems, metamaterials analysis and design, and antenna array synthesis. Prof. Oliveri served as an Associate Editor of the IEEE Antennas and Wireless Propagation Letters (2016-2022) and of the IEEE Journal on Multiscale and Multiphysics Computational Techniques (2017-2023), and he is AE of EPJ Applied Metamaterials, of the International Journal of Antennas and Propagation, of the International Journal of Distributed Sensor Networks, of the Microwave Processing journal, and of the Sensors journal. He has been serving as the Chair of the AP-S IEEE Press Liaison  Committee, as Member of the IEEE AP-S Field Award Subcommittee, and as Member of the IEEE AP-S Membership and Benefit Committee. He is the Chair of the IEEE AP/ED/MTT North Italy Chapter.\end{IEEEbiography}

\begin{IEEEbiography}[{\includegraphics[width=1in,height=1.25in,clip,keepaspectratio]{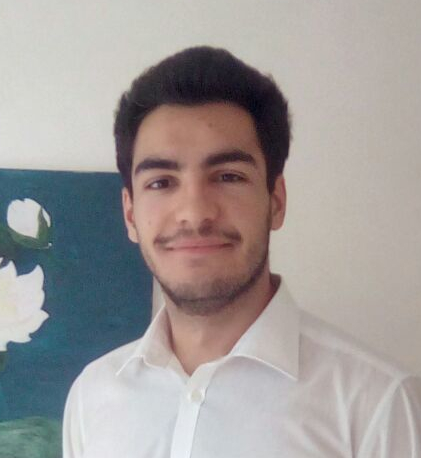}}]{Francesco Zardi} received the B.Sc. in Telecommunications and Electronic Engineering and the M.Sc. in Information and Communications Engineering from the University of Trento, Italy, in 2017 and 2019, respectively. He attended the International Doctoral School in Information and Communication Technology of Trento and is a Senior Researcher at the ELEDIA Research Center. His research activity is mainly focused on electromagnetic diagnostic techniques, the Smart Electromagnetic Environment, and advanced radar architectures.\end{IEEEbiography}

\begin{IEEEbiography}[{\includegraphics[width=1in,height=1.25in,clip,keepaspectratio]{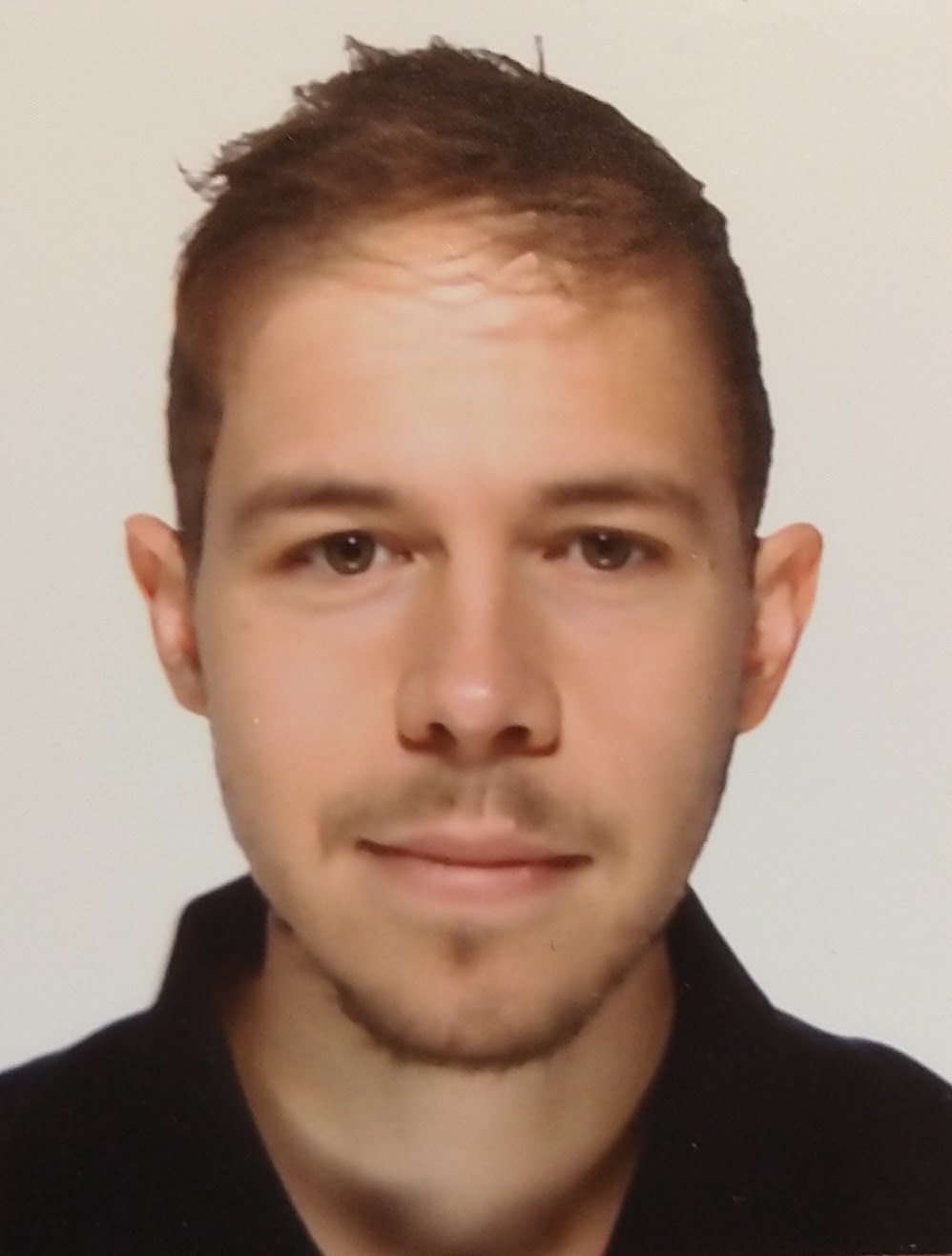}}]{Giorgio Gottardi} received the B.S. degree in Electronics and Telecommunication Engineering in 2012 and the M.S. degree in Telecommunication Engineering in 2015 at the University of Trento, Italy, and the PhD degree from the International Doctoral School in Information and Communication Technology of Trento in 2019. He is currently a PostDoc at the Department of Civil, Environmental and Mechanical Engineering (DICAM) at the University of Trento, and a Research Fellow of the ELEDIA Research Center. His research activities are mainly focused on synthesis methods for unconventional antenna array architectures for next generation communications.\end{IEEEbiography}

\begin{IEEEbiography}[{\includegraphics[width=1in,height=1.25in,clip,keepaspectratio]{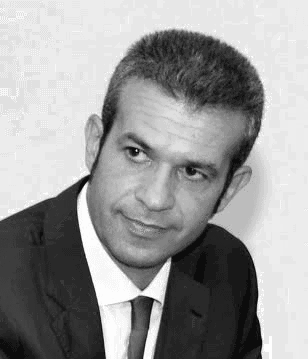}}]{Andrea Massa} (IEEE Fellow, IET Fellow, Electromagnetic Academy Fellow) received the Laurea (M.S.) degree in Electronic Engineering from the University of Genoa, Genoa, Italy, in 1992 and the Ph.D. degree in EECS from the same university in 1996. He is currently a Full Professor of Electromagnetic Fields at the University of Trento, where he currently teaches electromagnetic fields, inverse scattering techniques, antennas and wireless communications, wireless services and devices, and optimization techniques. At present, Prof. Massa is the director of the network of federated laboratories "ELEDIA Research Center" (www.eledia.org) [ELEDIA@UTB in Bandar Seri Begawan (Brunei), ELEDIA@TSINGHUA in Beijing (China), ELEDIA@UniCAS in Cassino (Italy), ELEDIA@UESTC in Chengdu (China), ElEDIA@UiC in Chicago (USA), ELEDIA@USIL in Lima (Peru), ELEDIA@UniNAGA in Nagasaki (Japan), ELEDIA@L2S in Paris (France), ELEDIA@CTU in Prague (Czech), ELEDIA@AUTH in Thessaloniki (Greece), ELEDIA@UniTN in Trento (Italy), ELEDIA@Innov'COM in Tunis (Tunisia), ELEDIA@XIDIAN in Xi'an (China)]. Moreover, he is holder of a Chang-Jiang Chair Professorship @ UESTC (Chengdu, China), Visiting Research Professor @ University of Illinois at Chicago (Chicago, USA), Visiting Professor @ Tsinghua (Beijing - China), Visiting Professor @ Tel Aviv University (Tel Aviv, Israel), and Professor @ CentraleSupelec (Paris - France). He has been holder of a Senior DIGITEO Chair at L2S-CentraleSupelec and CEA LIST in Saclay (France), UC3M-Santander Chair of Excellence @ Universidad Carlos III de Madrid (Spain), Adjunct Professor at Penn State University (USA), Guest Professor @ UESTC (China), and Visiting Professor at the Missouri University of Science and Technology (USA), the Nagasaki University (Japan), the University of Paris Sud (France), the Kumamoto University (Japan), and the National University of Singapore (Singapore). He has been appointed IEEE AP-SDiuished Lecture016-2018) and seved as Associate Editor of the "IEEE Transaction on Antennas and Propagation" (2011-2014).  Prof. Massa serves as Associate Editor of the "International Journal of Microwave and Wireless Technologies" and he is member of the Editorial Board of the "Journal of Electromagnetic Waves and Applications", a permanent member of the "PIERS Technical Committee" and of the "EuMW Technical Committee", and a ESoA member. He has been appointed in the Scientific Board of the "Societa  Italiana di Elettromagnetismo (SIEm)" and elected in the Scientific Board of the Interuniversity National Center for Telecommunications (CNIT). He has been appointed in 2011 by the National Agency for the Evaluation of the University System and National Research (ANVUR) as a member of the Recognized Expert Evaluation Group (Area 09, "Industrial and Information Engineering") for the evaluation of the researches at the Italian University and Research Center for the period 2004-2010. Furthermore, he has been elected as the Italian Member of the Management Committee of the COST Action TU1208 - Civil Engineering Applications of Ground Penetrating Radar. His research activities are mainly concerned with inverse problems, analysis/synthesis of antenna systems and large arrays, radar systems synthesis and signal processing, cross-layer optimization and planning of wireless/RF systems, semantic wireless technologies, system-by-design and material-by-design (metamaterials and reconfigurable-materials), and theory/applications of optimization techniques to engineering problems (tele-communications, medicine, and biology). Prof. Massa published more than 900 scientific publications among which more than 350 on international journals (> 14.700 citations, h-index = 63 [Scopus]; > 12.000 citations,  h-index = 58 [ISI-WoS]; > 23.900 citations, h-index = 88 [Google Scholar]) and more than 570 in international conferences where he presented more than 210 invited contributions (> 50 invited keynote speaker) (www.eledia.org/publicto.He ha rgnzed more than 100 scienific sessions in international conferences and has participated to several technological projects in the national and international framework with both national agencies and companies (18 international prj, > 5 MEu; 8 national prj, > 5 MEu; 10 local prj, > 2 MEu; 63 industrial prj, > 10 MEu; 6 university prj, > 300 KEu).\end{IEEEbiography}

\EOD
\end{document}